\documentclass[%
report,
nofootinbib,
 amsmath,amssymb,
 aps,
]{revtex4-1}

\usepackage{graphicx}
\usepackage{bm}
\usepackage{slashed}
\usepackage{color}

\begin{document}


\title{Black hole perturbation under $2+2$ decomposition in the action}

\author{Justin L. Ripley}
 \email{jripley@princeton.edu}
\author{Kent Yagi}%
 \email{kyagi@princeton.edu}
\affiliation{%
 Department of Physics, Princeton University, Princeton, New Jersey 08544, USA.
}%


\date{\today}

\begin{abstract}
	Black hole perturbation theory is useful for studying the stability of 
black holes and calculating ringdown gravitational waves after the 
collision of two black holes. Most previous calculations were carried 
out at the level of the field equations instead of the action. 
In this work, we compute the Einstein-Hilbert action to quadratic 
order in linear metric perturbations about a spherically symmetric 
vacuum background in Regge-Wheeler gauge. Using a $2+2$ splitting of spacetime,
we expand the metric perturbations into a sum over scalar, vector, and 
tensor spherical harmonics, and dimensionally reduce the action to two 
dimensions by integrating over the two sphere. We find that the axial 
perturbation degree of freedom is described by a two dimensional 
massive vector 
action, and that the polar perturbation degree of freedom is described 
by a two dimensional dilaton massive gravity action. 
Varying the dimensionally reduced actions, we rederive covariant
and gauge-invariant master equations for the axial and polar degrees
of freedom. 
Thus, the 
two dimensional massive vector and massive gravity actions we derive 
by dimensionally reducing the perturbed Einstein-Hilbert action
describe the dynamics of a well studied physical system: the metric 
perturbations of a static black hole. The $2+2$ formalism we present 
can be generalized 
to $m+n$ dimensional spacetime splittings, which may be useful in 
more generic situations, such as
expanding metric perturbations in higher dimensional gravity. We
provide a self-contained presentation of $m+n$ formalism for
vacuum spacetime splittings.

\end{abstract}

\maketitle


\section{\label{sec:Introduction}Introduction}
	The theory of metric perturbations of static black hole spacetimes is 
an old and well studied subject. The field began with the work of Regge 
and Wheeler~\cite{PhysRev.108.1063}, who were the first to study linear 
metric perturbations of the Schwarzschild background. In particular, 
Regge and Wheeler derived a closed form expression, the Regge-Wheeler 
equation, for linear axial perturbations. The field was further developed 
by several workers, most notably Vishveshwara~\cite{PhysRevD.1.2870} 
and Zerilli~\cite{PhysRevLett.24.737}, the latter of whom derived a closed 
form expression, the Zerilli equation, for linear polar perturbations. 
Black hole perturbation theory was first presented in a gauge invariant 
manner by Moncrief~\cite{MONCRIEF1974323}, who also wrote down a 
Hamiltonian for axial and polar perturbations. Gerlach and 
Sengupta~\cite{PhysRevD.19.2268,PhysRevD.22.1300} later formulated a 
covariant and gauge invariant formalism to describe 
the metric and matter perturbations of a generic spherically symmetric 
spacetime. A thorough exposition of the state of the field up until 
the mid 1980's can be found in Chandrasekhar's monograph on the 
subject~\cite{Chandrasekhar:579245}, while a more modern, covariant, 
and gauge invariant formulation of the theory of static black hole 
perturbations including source terms is presented by Martel and 
Poisson in~\cite{Martel:2005ir}. 

	While much progress has been made in understanding and reformulating 
the equations of motion of metric perturbations of static black holes, 
less work has been done on understanding the structure of the perturbed 
Einstein-Hilbert action in this background (see, however 
~\cite{Kobayashi:2012kh,Kobayashi:2014wsa,DeFelice:2011ka,Motohashi:2011pw,
Motohashi:2011ds,Ogawa:2015pea,Takahashi:2015pad,Takahashi:2016dnv}). 
The purpose of this 
paper is to further develop this aspect of metric perturbation theory. 
There are several reasons why deriving the action for black hole 
perturbations may be useful, two of which we briefly describe below. 
	
	Firstly, this formalism may be useful in constructing effective 
field theories of black hole ringdown. In the context of FLRW cosmologies, 
a $1+3$ decomposition is natural as the background can be naturally split 
into a three dimensional maximally symmetric spacelike hypersurface and a 
time direction. The maximally symmetric subspace of a Schwarzschild
black hole is the two sphere. Adapting an effective field 
theory approach with a $2+2$ formalism may be more 
useful for this background, where a foliation by maximally
symmetric subspaces would be by two dimensional spheres
(see, for example~\cite{Kase:2014baa} for a related discussion). 

	Secondly, deriving the action for black hole perturbations may also be 
useful in understanding the quantum physics of black holes. Two dimensional 
gravity has been used to study Hawking radiation and the quantum mechanics 
of black holes~\cite{PhysRevD.14.870,1993stqg.conf..122H,
Vagenas:2001sm}. The actions in Eqs.~\eqref{Saxial} 
and~\eqref{Spolar} could be useful in this context; for example in the 
construction of a path integral formulation of Hawking radiation 
for the metric perturbations of a black hole.  
 
	In this article, we derive the perturbed Einstein-Hilbert action for 
spherically symmetric backgrounds. From this action, we derive the 
equations of motion for the Schwarzschild black hole. We derive the action 
using a $2+2$ spacetime splitting, which allows us to decouple the linear 
scalar, vector, and tensor (SVT) perturbations in the action. While this 
is not the first work that derives black hole perturbations from the action
~\cite{Kobayashi:2012kh,Kobayashi:2014wsa,DeFelice:2011ka,Motohashi:2011pw,
Motohashi:2011ds,Ogawa:2015pea,Takahashi:2015pad,Takahashi:2016dnv}, to our 
knowledge the application of the $2+2$ formalism directly to the perturbed 
Einstein-Hilbert action is novel, and brings to light several interesting 
new conceptual points about the nature of black hole perturbations. For 
example, we find that the polar perturbations of a Schwarzschild black hole 
are described by a (1+1)-dimensional dilaton massive gravity model, which 
naturally emerges by dimensionally reducing the perturbed Einstein-Hilbert 
action in a spherically symmetric background. Additionally, the axial 
perturbations of a black hole are described by a (1+1)-dimensional 
massive vector field action. While we derive these actions
in the Regge-Wheeler gauge, our results may trivially be reexpressed
in a gauge invariant fashion (see, for example, section IV B of
\cite{Martel:2005ir}).  

	The organization of this paper is as follows.
	In Sec.~\ref{sec:EinsteinHilbertnpm}, we briefly review the $m+n$ 
formalism for vacuum spacetime splittings  
as applied to the Einstein-Hilbert action; more details are given in 
Appendices~\ref{generalcoDimensionGeometry} 
and \ref{ADMLikeVariables}. In Sec.~\ref{LinearPert}, we set our notation 
and review metric perturbation theory in a spherically symmetric spacetime. 
In Secs.~\ref{AxialPert} and~\ref{PolarPert} we derive the action for axial 
and polar perturbations respectively for a 
spherically symmetric background, which we derive in the Regge-Wheeler 
gauge~\cite{PhysRev.108.1063}. From the axial and polar equations of motion,
we rederive covariant and gauge invariant expressions for the axial
and polar degrees of freedom, respectively.
We discuss our results and 
conclude in Sec.~\ref{sec:conclusion}. We review the mathematics of 
the geometry of surfaces of arbitrary codimension in 
Appendix~\ref{generalcoDimensionGeometry}, compute the Einstein-Hilbert action
in ADM-like variables adapted to higher codimension spacetime splittings
in Appendix~\ref{ADMLikeVariables}, and provide a summary of the properties of 
scalar, vector, and tensor spherical harmonics in Appendix~\ref{SVTSpherical}.

	Our sign conventions for the metric and Riemann tensor follow that of 
Misner, Thorne, and Wheeler~\cite{misner1973gravitation}: for a Lorentzian
manifold the metric 
signature is $-++\cdots$, and $R^{\alpha}{}_{\mu\beta\nu} = 
\partial_{\beta}\Gamma^{\alpha}_{\mu\nu} - \cdots$. We work in reduced 
Planck units: $8\pi G = c = \cdots = 1$.  

\section{Einstein-Hilbert action in the $m+n$ formalism
\label{sec:EinsteinHilbertnpm}}
	In this section we briefly review the $m+n$ formalism for vacuum spacetime
splittings. A more 
detailed description of this formalism is presented in 
Appendices~\ref{generalcoDimensionGeometry} and~\ref{ADMLikeVariables}. 

	We begin with a $d$ dimensional (semi-)Riemannian manifold $M$
with metric $g_{\mu\nu}$ and connection $\nabla_{\mu}$. We assume that
the topology of $M$ is $\mathbb{R}^m\times\Sigma$, so that we may foliate
$M$ with a family of $n=d-m$ dimensional submanifolds
$\{\Sigma_{\bf t}\}_{\bf t\in\mathbb{R}^m}$. Unless otherwise noted
we  will drop the index subscript ${\bf t}$ from $\Sigma_{\bf t}$.  
For every point 
$p\in\Sigma$, the tangent space of $p$ 
naturally splits into a tangent
and transverse space, $T_p(M) = T_p(\Sigma)\oplus T_p({}^{\perp}\Sigma)$.
We define the tangent projector 
on to the tangent space $h_{\mu}{}^{\nu}$ and the
transverse projector $l_{\mu}{}^{\nu} = \delta_{\mu}{}^{\nu}-h_{\mu}{}^{\nu}$. 

Let us define the notion of tangent and transverse in more detail.
	A tensor \emph{component} is called tangent if its contraction with
the transverse projector is zero; e.g. if $l_{\mu}{}^{\nu}P^{\mu\alpha}=0$ 
then we say the $\mu$ component of $P^{\mu\alpha}$ is tangent. Likewise
a component of a tensor is called transverse if its contraction
with the tangent projector is zero. A tensor is called 
tangent (transverse) if all of its components are tangent (transverse).
We define the tangent extrinsic curvature $K^{\gamma}{}_{\mu\nu}$ and
the transverse extrinsic curvature $A^{\gamma}{}_{\mu\nu}$
\begin{align}
	K^{\gamma}{}_{\mu\nu} 
	\equiv & 
	h_{\mu}{}^{\alpha}h_{\nu}{}^{\beta}\nabla_{\beta}l_{\alpha}{}^{\gamma}\,,
	\\ 
	A^{\gamma}{}_{\mu\nu}
	\equiv & 
	l_{\mu}{}^{\alpha}l_{\nu}{}^{\beta}\nabla_{\beta}h_{\alpha}{}^{\gamma}\,.
\end{align} 
	We define a tangent 
derivative operator ${}^{\parallel}\nabla_{\mu}$ as the
tangent projection
 of the action of $\nabla_{\mu}$ on a tangent tensor; e.g.
for some $v_{\mu}\in T_p^*(\Sigma)$ we would have
${}^{\parallel}\nabla_{\mu}v_{\nu}
= h_{\mu}{}^{\alpha}h_{\nu}{}^{\beta}\nabla_{\alpha}v_{\beta}$.
Likewise we define the transverse derivative operator ${}^{\perp}\nabla_{\mu}$
as the transverse projection the action of $\nabla_{\mu}$ on a
transverse tensor. As the transverse spaces will generally not integrate
to form a set of submanifolds, the transverse derivative will generally not be
torsion free. We define curvature tensors for the tangent and transverse
tensors as follows. For any $v_{\mu}\in T^*_p(\Sigma)$, we define
\begin{equation}
	{}^{\parallel}R_{\alpha\beta\gamma}{}^{\delta}v_{\delta}
	\equiv 2{}^{\parallel}\nabla_{[\alpha}{}^{\parallel}\nabla_{\beta]}v_{\gamma}\,.	
\end{equation}
	Similarly, for any $v_{\mu}\in T^*_p({}^{\perp}\Sigma)$, we define
\begin{equation}
	{}^{\perp}R_{\alpha\beta\gamma}{}^{\delta}v_{\delta}
	\equiv 2{}^{\perp}\nabla_{[\alpha}{}^{\perp}\nabla_{\beta]}v_{\gamma}
	+ F^{\lambda}{}_{\alpha\beta}l_{\gamma}{}^{\delta}
	\nabla_{\lambda}v_{\delta}	\,,
\end{equation}
where the transverse torsion tensor is defined by
$F^{\gamma}{}_{\alpha\beta} \equiv 2A^{\gamma}{}_{[\alpha\beta]}$. 
	With these definitions at hand, we can rewrite the Ricci scalar as follows
\begin{align}
	R = & \left(
	h^{\alpha\gamma}h^{\beta\delta} + l^{\alpha\gamma}l^{\beta\delta}
	+ 2 h^{\alpha\gamma}l^{\beta\delta}
		\right) 
	R_{\alpha\beta\gamma\delta}  \nonumber \\
	= & {}^{\parallel}R 
	  + {}^{\perp}R 
	+ K_{\lambda}K^{\lambda}		
	- K_{\lambda\alpha\beta}K^{\lambda\beta\alpha}
	+ A_{\lambda}A^{\lambda}		
	- A_{\lambda\alpha\beta}A^{\lambda\beta\alpha}
	  - 2 \nabla_{\lambda}\left(K^{\lambda} 
			+ A^{\lambda}\right) , 
\end{align}
where $K^{\lambda} \equiv K^{\lambda}{}_\mu{}^\mu$ and 
$A^{\lambda} \equiv A^{\lambda}{}_\mu{}^\mu$.

	At this point we choose a basis adapted to the $m+n$ foliation. 
Our discussion here most closely follows that of \cite{Brady:1995na}. The
coordinates $\{x^{\alpha}\}$ of some chart of the spacetime manifold $M$
are written as functions of two sets of variables, $\{u^a\}$ and
$\{\theta^A\}$, so $x^{\alpha}\equiv x^{\alpha}(u^a,\theta^A)$. Our notation
is as follows: Greek indices run from $0,...,d-1$, lower case Latin indices
run from $0,...,m-1$ and upper case Latin indices from from $m,...,d-1$. 
Einstein summation notation will apply to all different index types. 
Derivatives with respect to the variables $\{u^a\}$ will be denoted
by $\partial_a\equiv\partial/\partial u^a$, while derivatives with respect
to the variables $\{\theta^A\}$ will be denoted by 
$\partial_A\equiv\partial/\partial\theta^A$. We set the variables
$\{\theta^A\}$ to be intrinsic to the leaf $\Sigma$. We define
a basis of frame vectors $e^{\alpha}_A\equiv\partial_Ax^{\alpha}$
which span $T_p(\Sigma)$. The first fundamental form of $\Sigma$
is $\gamma_{AB} = g_{\alpha\beta}e^{\alpha}_Ae^{\beta}_B$; the inverse
of $\gamma_{AB}$ is $\gamma^{AB}$, and the 
metric compatible induced covariant derivative is denoted by
${}^{\parallel}\nabla_A$. Upper case Latin indices are raised/lowered
by $\gamma^{AB}$ and $\gamma_{AB}$, respectively.
The variables $\{u^a\}$, which may also be
thought of as functions on the chart,
are constant on each leaf. We define a congruence of vector fields
$u^{\gamma}_c\equiv\partial_cx^{\gamma}$ upon which the frame
vectors $\{e^{\alpha}_A\}$ are Lie transported. We next define a basis for 
$T_p^*({}^{\perp}\Sigma)$, $n^a_{\alpha}\equiv\partial_{\alpha}u^a$. The 
components of the inner product matrix of the forms $n^a_{\alpha}$ 
is written as $\alpha^{ab}=g^{\alpha\beta}n^a_{\alpha}n^b_{\beta}$.
The matrix inverse of $\alpha^{ab}$ is denoted by $\alpha_{ab}$. Formally,
we will raise/lower lower case Latin indices with $\alpha^{ab}$ and
$\alpha_{ab}$, respectively. We note that generally $\alpha_{ab}$ is
generically \emph{not} the first fundamental form of any submanifold as 
the transverse spaces generally do not integrate to form a submanifold. 
We decompose
the differential $dx^{\alpha}$ into terms tangent and transverse to
the leaf $\Sigma$,
\begin{equation}
	dx^{\alpha} 
	= n^{\alpha}_adu^a 
	+ e^{\alpha}_A\left(d\theta^A + \beta^A_adu^a\right) ,
\end{equation} 
	where we have defined the shift vectors $\{\beta^{\alpha}_a\}$. We
now write down the line element for this adapted basis
\begin{equation}
	ds^2
	= 
	\alpha_{ab}du^adu^b 
	+ \gamma_{AB}
	\left(d\theta^A + \beta^A_adu^a\right)
	\left(d\theta^B + \beta^B_bdu^b\right) ,
\end{equation} 
where we recall $g_{\alpha \beta} e_A^\alpha n_b^\beta=0$. 
	With this line element the metric determinant factorizes as 
follows: $\mathrm{det}g = \mathrm{det}\alpha\, \mathrm{det}\gamma$.
	We can now compute the curvatures 
$K^{\gamma}{}_{\alpha\beta}$, 
$A^{\gamma}{}_{\alpha\beta}$, 
${}^{\parallel}R_{\alpha\beta\gamma\delta}$, and
${}^{\perp}R_{\alpha\beta\gamma\delta}$ in terms of the 
metric components $\alpha_{ab}$, $\gamma_{AB}$, and $\beta^A_a$:
\allowdisplaybreaks
\begin{align}
	K_{\gamma\alpha\beta} 
	= & 
	e^A_{\alpha}e^B_{\beta}n^c_{\gamma}\mathcal{K}_{cAB}
	, \\
	A_{\gamma\alpha\beta}
	= & 
	e_{C\gamma}n^a_{\alpha}n^b_{\beta}\mathcal{A}^C_{ab}
	, \\
	{}^{\parallel}R_{\alpha\beta\gamma\delta}
	= & 
	e^A_{\alpha}e^B_{\beta}e^C_{\gamma}e^D_{\delta}
	{}^{\parallel}\mathcal{R}_{ABCD}
	, \\
	{}^{\perp}R_{\alpha\beta\gamma\delta}
	= & 
	n^a_{\alpha}n^b_{\beta}n^c_{\gamma}n^d_{\delta}
	\left(
	{}^{\perp}\mathcal{R}_{abcd}
	+ 2\alpha_{ai}\alpha_{bj}\alpha_{ck}\alpha_{dl}\gamma^{CD}
	\mathcal{A}^{[ij]}_C\mathcal{A}^{lk}_D
	\right) 
	, 
\end{align}
	where 
\begin{align}
	\mathcal{K}_{cAB} 
	= & 
	\frac{1}{2}
	\left(
	\partial_c\gamma_{AB}
	- {}^{\parallel}\nabla_A\beta_{cB}
	- {}^{\parallel}\nabla_B\beta_{cA}
	\right)
	, \\
	\mathcal{A}^{ab}_C
	= & 
	\frac{1}{2}
	\left(
	\partial_C\alpha^{ab}	
	- \alpha^{ac}\alpha^{bd}\gamma_{CD}\mathcal{F}^D_{cd}
	\right)
	, \\
	\mathcal{F}^C_{ab}
	= & 
	  \partial_a\beta^C_b - \partial_b\beta^C_a 
	+ \beta^D_b\partial_D\beta^C_a - \beta^D_a\partial_D\beta^C_b
	, \\
	{}^{\parallel}\mathcal{R}^D{}_{CAB}
	= & 
	  \partial_A\Gamma^D{}_{CB} - \partial_B\Gamma^D{}_{CA} 
	+ \Gamma^D{}_{IA}\Gamma{}^I{}_{CB} - \Gamma^D{}_{IB}\Gamma^I{}_{CA}
	, \\
	\Gamma_{CAB}
	= & 
	\frac{1}{2}
	\left(
	\partial_A\gamma_{CB} + \partial_B\gamma_{CA} - \partial_C\gamma_{AB}
	\right)
	, \\
	{}^{\perp}\mathcal{R}^d{}_{cab}
	= & 
	  n_a^{\mu}\partial_{\mu}\Omega^d{}_{cb} 
	- n_b^{\mu}\partial_{\mu}\Omega^d{}_{ca}
	+ \Omega^d{}_{ia}\Omega^i{}_{cb} - \Omega^d{}_{ib}\Omega^i{}_{ca}
	, \\
	\Omega_{cab}
	= &
	\frac{1}{2}
	\left(
	n^{\mu}_a\partial_{\mu}\alpha_{cb} + n^{\mu}_b\partial_{\mu}\alpha_{ca}
	- n^{\mu}_c\partial_{\mu}\alpha_{ab}
	\right) . 
\end{align}
	The Einstein-Hilbert action in this formalism can be written as 
\begin{align}
\label{eq:EHActionADM_Sec1}
	S_{EH} = \int d^mud^n\theta 
	\sqrt{\alpha}\sqrt{\gamma}
	\Big( &  
	{}^{\parallel}\mathcal{R}
	+ \alpha^{cd}\gamma^{AB}\gamma^{CD}
		\left(
	  \mathcal{K}_{cAB}\mathcal{K}_{dCD} 
	- \mathcal{K}_{cAC}\mathcal{K}_{dBD} 
		\right)
	\nonumber \\ &  
	+ {}^{\perp}\mathcal{R}
	+ \gamma^{CD}\alpha_{ab}\alpha_{cd}
		\left(
	  \mathcal{A}_C^{ab}\mathcal{A}_D^{cd} 
	- \mathcal{A}_C^{ac}\mathcal{A}_D^{bd} 
		\right)
	- 2 \nabla_{\lambda}\left(K^{\lambda}+A^{\lambda}\right)
	\Big) .
\end{align}
	We direct the reader to Appendices~\ref{generalcoDimensionGeometry} 
and~\ref{ADMLikeVariables} for a more detailed
discussion of the $m+n$ formalism, including a discussion of the relation
of this formalism to the ADM $1+(d-1)$ formalism,
and for derivations of the main results stated in this section.

\section{\label{LinearPert}
Metric perturbations for spherically symmetric background}

	In this section, we consider perturbations around a spherically 
symmetric four dimensional background spacetime. 
In a spherically symmetric spacetime, 
the full spacetime manifold
naturally factorizes into the form $M=M^2\times S^2$, where both 
$M^2$ and $S^2$ are submanifolds of $M$. $S^2$ is the two-sphere and roughly 
speaking $M^2$ is the `(t,r) plane' (see, for example the discussion
in section II of \cite{Martel:2005ir}). 
For factorizable spacetimes the metric naturally factorizes as well; i.e.
we can choose a background metric such that the shift vectors
$\{\beta^{\alpha}_a\}$ are all zero. 

We write the background metric as
\begin{equation}
\label{bkgrdSpherSym}
	ds^2 = {}^{(0)}\alpha_{ab}du^adu^b + {}^{(0)}\gamma_{AB}d\theta^Ad\theta^B . 
\end{equation}
We identify ${}^{(0)}\alpha_{ab}$ and ${}^{(0)}\gamma_{AB}$ as the 
metrics for $M^2$ and $S^2$, respectively. The metric 
${}^{(0)}\gamma_{AB}$ is equal to $r^2\Omega_{AB}$, where $\Omega_{AB}$ 
is the round metric. 
For a factorizable spacetime and metric
we may also interpret ${}^{(0)}\alpha_{ab}$ as the induced metric on $M^2$, 
and define
a metric compatible covariant derivative ${}^{\perp}{\nabla}_a$, with
$\Omega_{cab}$ as the connection coefficients. See Appendix 
\ref{factorizableSpacetime}
for a discussion of the $m+n$ formalism and factorizable spacetimes.

	We begin by describing the geometry of a linearly perturbed spherically 
symmetric background. We write	
\begin{align} 
\label{alpha_pert}
	\alpha_{ab} = & {}^{(0)}\alpha_{ab} + \delta\alpha_{ab} , \\
\label{beta_pert}
	\beta^A_a = & \delta \beta^A_a , \\
\label{gamma_pert}
	\gamma_{AB} = & {}^{(0)}\gamma_{AB} + \delta\gamma_{AB} .
\end{align}
	The perturbations $\delta\alpha_{ab}$, $\delta \beta^A_a$, and 
$\delta\gamma_{AB}$ can be split into pieces that transform as scalars, 
vectors, and tensors with respect to the $SO(3)$ spacetime isometry. This 
is accomplished by decomposing $\delta\alpha_{ab}$, 
$\delta\beta_{aA}={}^{(0)}\gamma_{AB}\delta\beta^B_a$, and 
$\delta\gamma_{AB}$ into a sum over spherical harmonics as
\begin{align}
\label{habSphHarm}
	\delta\alpha_{ab}(u^a,\theta^A) = & 
	\sum_{l,m}  h_{ab}^{lm}(u^a)Y^{lm}(\theta^A) , \\
\label{NAaSphHarm}
	\delta\beta_{Aa}(u^a,\theta^A) = & 
	r^2\sum_{l,m}\left\{ j^{lm}_a(u^a) E^{lm}_A(\theta^A) 
	+ h^{lm}_a(u^a)B^{lm}_A(\theta^A) \right\} , \\
\label{gamABSphHarm}
	\gamma_{AB} + \delta\gamma_{AB}(u^a,\theta^A) = & r^2 \sum_{l,m} 
\Big\{ \mathrm{exp}\left[2 k^{lm}(u^a)\right]\Omega_{AB}Y^{lm}(\theta^A) 
 \nonumber \\
& \qquad\qquad + G^{lm}(u^a)E^{lm}_{AB}(\theta^A) 
+ h_2^{lm}(u^a)B^{lm}_{AB}(\theta^A) \Big\} , 
\end{align}
	where $Y^{lm}$, $\left\{E^{lm}_A,B^{lm}_A\right\}$, and 
$\left\{E^{lm}_{AB},B^{lm}_{AB}\right\}$ are scalar, vector, and tensor 
spherical harmonics, respectively. We collect the basic properties of these 
functions in Appendix~\ref{SVTSpherical}. Our notation for the 
spherical harmonic decomposed perturbations follows Poisson and Martel
\cite{Martel:2005ir}, with the exceptions of $K^{lm}$, which we set to 
be $K^{lm}\equiv e^{2 k^{lm}}-1$ (see their equation (4.3)), and the 
perturbations $j_a^{lm}$ and $h_a^{lm}$, which we multiply by $r^2$
(see their equations (4.2) and (5.2)). 
We further note that unlike Martel and Poisson \cite{Martel:2005ir},
we raise/lower in indices $A$ with $\gamma_{AB}$, and 
\emph{not} the round metric $\Omega_{AB}$. This includes the indices of
the vector and tensor spherical harmonics.
With the decomposition in 
Eqs.~\eqref{habSphHarm}--\eqref{gamABSphHarm}, we have rewritten the ten 
metric perturbation degrees of freedom into a sum over SVT spherical 
harmonics. We see that there are four scalar, four vector, and two tensor 
spherical harmonic degrees of freedom. In reduced Planck units the variables 
$\left\{ h_{ab}^{lm},  k^{lm}, G^{lm}, h_2^{lm}\right\}$ 
are dimensionless, while the variables $\left\{j^{lm}_a, h^{lm}_a\right\}$ 
have dimensions of inverse length. 
 
	For completeness, 
we next review the gauge transformations of the perturbed quantities. Our
treatment and notation most closely follows that of Martel and Poisson
\cite{Martel:2005ir}. 
A linear gauge transformation can be written as the Lie derivative of the 
background metric along some arbitrary infinitesimal vector $\xi^{\mu}$:
\begin{equation}
\label{gaugeTransformation}
	\pounds_{\xi^{\alpha}}g_{\mu\nu} = \xi^{\alpha}\partial_{\alpha}g_{\mu\nu}
+ g_{\mu\alpha}\partial_{\nu}\xi^{\alpha} 
+ g_{\nu\alpha}\partial_{\mu}\xi^{\alpha} 
= \nabla_{\mu}\xi_{\nu} + \nabla_{\nu}\xi_{\mu} . 
\end{equation}
	Under these transformations, and with our line element in 
Eqs.~\eqref{bkgrdSpherSym},~\eqref{alpha_pert},~\eqref{beta_pert},
and \eqref{gamma_pert}, we see that our perturbations 
$\delta\alpha_{ab}$, $\delta\beta_a^A$, and $\delta\gamma_{AB}$ transform as
\begin{equation}
\begin{aligned}
\label{SphericalGaugeTrans}
	\delta\alpha_{ab} \to & \delta\alpha_{ab} 
	+ {}^{\perp}\nabla_a\xi_b + {}^{\perp}\nabla_b\xi_a 
	, \\
	\delta\beta_{aA} \to & \delta\beta_{aA} + {}^{(0)}\alpha_{ac}\partial_A\xi^c 
	+ r^2\Omega_{AC}\partial_a\xi^C , \\
	\delta\gamma_{AB} \to & \delta\gamma_{AB} + \Omega_{AB}\xi^c\partial_cr^2
	+ {}^{\parallel}\nabla_A\xi_B + {}^{\parallel}\nabla_B\xi_A . 
\end{aligned}
\end{equation}
	We can split the four-vector $\xi^{\mu}$ into terms that transform as 
scalars and vectors with respect to the $SO(3)$ isometry:
\begin{equation}
\begin{aligned}
\label{xiDecomp}
	\xi_a(x^a,\theta^A) = & \sum_{l,m} (\xi^{lm}_{\mathcal{S}})_a(u^a)\, 
	Y^{lm}(\theta^A) , \\
	\xi_A(x^a,\theta^A) = & r^2 \sum_{l,m} 
	\left\{
	  \xi_{\mathcal{E}}^{lm}(u^a)\,E^{lm}_A(\theta^A) 
	+ \xi_{\mathcal{B}}^{lm}(u^a)\, B^{lm}_A(\theta^A) \right\} , 
\end{aligned}
\end{equation}
	where the label $\mathcal{S}$, stands for `scalar part', $\mathcal{E}$ 
for `electric (polar)' part, and $\mathcal{B}$ for `magnetic (axial)' part
of the black hole perturbations. 
We see that $\xi_{\mu}$ 
has two scalar and two vector degree of freedom, one of which is axial and 
the other which is polar. Note that we have chosen to normalize the scalars 
and vectors so that in reduced Planck units the quantities
$\left\{(\xi^{lm}_{\mathcal{S}})_a\right\}$ have the dimension of 
length, while quantities $\left\{\xi^{lm}_{\mathcal{E}}, 
\xi^{lm}_{\mathcal{B}}\right\}$ 
are dimensionless. 
In Table~\ref{SVTgaugeTransformations} we list how the SVT components of 
$\delta\alpha_{ab}$, $\beta^A_a$, and $\delta\gamma_{AB}$ transform 
under the gauge 
transformation in Eq.~\eqref{gaugeTransformation}. 
\begin{table}
\caption{Gauge transformations for spherically symmetric background given 
by Eq.~\eqref{bkgrdSpherSym}.}
\begin{center}
\label{SVTgaugeTransformations}
\begin{tabular}{ c | c | c}
\hline
	& variable(s) & gauge transformation \\	
\hline
	scalar & $ h_{ab}^{lm}$ & $  h_{ab}^{lm}\to 
	 h_{ab}^{lm} 
	+ {}^{\perp}\nabla_a(\xi^{lm}_{\mathcal{S}})_b 
	+ {}^{\perp}\nabla_b(\xi^{lm}_{\mathcal{S}})_a$ \\
	& $ k^{lm} $ & $ k^{lm} \to  k^{lm} 
+ \frac{1}{2} \frac{1}{r^2} (\xi^{lm}_{\mathcal{S}})^a\partial_ar^2 
- \frac{1}{2}l(l+1)\xi^{lm}_{\mathcal{E}}$ \\
\hline
	vector & $j_a^{lm} $ & $j_a^{lm}\to j^{lm}_a 
+ \frac{1}{r^2}(\xi^{lm}_{\mathcal{S}})_a + \partial_a \xi^{lm}_{\mathcal{E}}$ \\
	& $h_a^{lm} $ & $h_a^{lm}\to h_a^{lm} + \partial_a\xi^{lm}_{\mathcal{B}}$ \\
\hline
	tensor & $G^{lm} $ & $G^{lm}\to G^{lm} + 2\xi^{lm}_{\mathcal{E}}$ \\
	& $h_2^{lm} $ & $h_2^{lm}\to h_2^{lm} + 2 \xi^{lm}_{\mathcal{B}}$ \\ 
\hline
\end{tabular} 
\end{center}
\end{table}
	Unlike in cosmological perturbation theory~\cite{PhysRevD.22.1882}, 
the tensor perturbations with respect to the 
(spherically symmetric) background are not 
gauge invariant. Using the relations listed in 
Table~\ref{SVTgaugeTransformations}, one can construct gauge-invariant 
perturbations~\cite{MONCRIEF1974323,Martel:2005ir}, which we list
for completeness 
\begin{align}
	\tilde{h}_a^{lm} \equiv & h_a^{lm} - \frac{1}{2}\partial_ah_2^{lm} , \\
	\tilde{h}^{lm}_{ab} \equiv &  h_{ab}^{lm} 
	- {}^{\perp}\nabla_a\epsilon_b - {}^{\perp}\nabla_b\epsilon_a , \\
	\tilde{k}^{lm} \equiv &  k^{lm} - \frac{1}{2}\epsilon^a\partial_ar^2 
	+ \frac{1}{4}l(l+1)G^{lm} ,	
\end{align} 
	where $\epsilon_a$ is defined to be \cite{Martel:2005ir} 
\begin{align} 
	\epsilon_a = r^2 j^{lm}_a - \frac{1}{2}r^2\partial_aG^{lm} .
\end{align}
	We see that $\tilde{h}^{lm}_a$ is an axial, while $\tilde{h}^{lm}_{ab}$ 
and $\tilde{k}^{lm}$ are polar gauge invariant perturbation variables.
		
	In this paper, we adopt the Regge-Wheeler gauge. Such a gauge fixes 
the scalar and vector components of the gauge vector $\xi_{\mu}^{lm}$ as follows:
\begin{align}
\label{xiB}
	\xi^{lm}_{\mathcal{B}} = & - \frac{1}{2}h_2^{lm} , \\
\label{xiE}
	\xi^{lm}_{\mathcal{E}} = & - \frac{1}{2}G^{lm} , \\
\label{xia}
	\left(\xi^{lm}_{\mathcal{S}}\right)_a = 
	& - r^2\partial_a \xi^{lm}_{\mathcal{E}} - r^2 j^{lm}_a . 
\end{align}
	While Regge and Wheeler worked with Schwarzschild coordinates
~\cite{PhysRev.108.1063}, we see that their gauge choice does not depend on 
the detailed structure of the two-metric $^{(0)}\alpha_{ab}$, insofar that it 
has no functional dependence on the angular variables  
$\{\theta^A\}$~\cite{Martel:2005ir}. 

	Importantly, as the gauge vector $\xi_{\mu}^{lm}$ is 
uniquely determined (e.g. with no integration constants)
by the conditions in Eqs.~\eqref{xiB}--\eqref{xia},  
we can derive the correct perturbation and background equations of motion by 
imposing the gauge conditions first and then varying the expanded 
Einstein-Hilbert action~\cite{Motohashi:2016prk}. The Regge-Wheeler gauge 
leaves us with the following six (two vector, four scalar) degrees of 
freedom: $\{h^{lm}_a,  h_{ab}^{lm},  k^{lm}\}$.

	Only one scalar and one vector degree of freedom, which correspond to 
the two polarizations of a gravitational wave, are dynamical degrees of 
freedom. The other three scalar degrees of freedom are either fixed by the 
equations of motion to be constants, or are absorbed into the definition 
of the Zerilli function $\Psi^{lm}_{even}$, which describes the dynamics of 
the polar perturbation
~\cite{PhysRevLett.24.737,MONCRIEF1974323,Martel:2005ir}.  
For the remainder of this paper all of our calculations will be 
performed in the Regge-Wheeler gauge. 
From the gauge transformations
listed in Table~\ref{SVTgaugeTransformations}, 
we see that we can rewrite our formulas in terms of the
gauge invariant variables using the relations as follows:
$h_a^{lm}\to\tilde{h}_a^{lm}$, $h_{ab}^{lm}\to\tilde{h}_{ab}^{lm}$,
and $k^{lm}\to\tilde{k}^{lm}$, so that all the formulas we list 
can be cast into a gauge invariant form (see for example 
\cite{Martel:2005ir}).

\section{\label{PertEHRWG}
Perturbed Einstein-Hilbert action in Regge-Wheeler gauge}

	In this section, we consider axial and polar perturbations of the 
Einstein-Hilbert action in the Regge-Wheeler gauge.	

\subsection{\label{bkgrdEOM}Background equations of motion}
	For completeness, we first derive the background equations of motion
from unperturbed Einstein-Hilbert action in spherical symmetry. The
unperturbed dimensionally reduced action is
\begin{align}
\label{EHAction_unperturbed}
	S = \int d^2u\sqrt{\alpha}
	\left(
	\frac{r^2}{2}{}^{\perp}\mathcal{R}
	+ (\partial_ar)^2 + 1
	\right)
	.
\end{align}
	Varying $r$ and $\alpha^{ab}$, we obtain the standard (see for example
appendix B of 
\cite{frolov2012black}) equations of motion
\begin{align}
\label{deltarbkgrdEOM}
	0 = & 
	r {}^{\perp}\mathcal{R} - 2 {}^{\perp}\Box r
	, \\
\label{deltaalphabkgrdEOM}
	0 = & 
	\left(2r{}^{\perp}\Box r + (\partial_cr)^2-1\right)\alpha_{ab}
	- 2r{}^{\perp}\nabla_a{}^{\perp}\nabla_br
	.	
\end{align}
	We note that we can split up Eq.~\eqref{deltaalphabkgrdEOM} 
by computing its trace and trace free components. The trace
gives us $r{}^{\perp}\Box r+(\partial r)^2 - 1=0$. We then use this in
Eq.~\eqref{deltaalphabkgrdEOM} to obtain
${}^{\perp}\nabla_a{}^{\perp}\nabla_b r 
= \frac{1}{2}\alpha_{ab}{}^{\perp}\Box r$ (see, for example Eq. (2.8) of
\cite{Martel:2005ir} for a similar expression). 

\subsection{\label{AxialPert}Axial perturbations}

\subsubsection{Axial action}
	Let us first consider axial perturbations. In the Regge-Wheeler gauge, 
the nonzero axial perturbations are completely described by the 
variable $h_a^{lm}$:
\begin{equation}
\label{RWAxial}
	\delta\beta_{Aa} = r^2\sum_{l,m} h^{lm}_a B^{lm}_A ,
\end{equation}
	in other words we consider the line element
\begin{equation}
\label{eq:RWAxialMetric}
	ds^2 = {}^{(0)}\alpha_{ab}du^adu^b 
	+ {}^{(0)}\gamma_{AB}
	\left(d\theta^A+\delta\beta^A_adu^a\right)
	\left(d\theta^B+\delta\beta^B_bdu^b\right) 
	,
\end{equation}
	with $\delta\beta^A_a=\gamma^{AB}\delta\beta_{aB}$ given by 
Eq.~\eqref{RWAxial}. For the remainder of this subsection 
our notation will be ${}^{(0)}\alpha_{ab}\equiv\alpha_{ab}$. 
The Einstein-Hilbert action expanded to  
linear order in $h^{lm}_a$ is 
zero in Regge-Wheeler gauge. So, we only need to consider the action expanded 
to quadratic order in $h_a^{lm}$. The terms of the Einstein-Hilbert action, 
 Eq.~\eqref{eq:EHActionADM_Sec1}, 
that are nonzero with line element Eq.~\eqref{eq:RWAxialMetric} are 
\begin{align}
\label{eq:S2odd}
	S^{(2)}_{odd} = \int d^4x \sqrt{\alpha}\sqrt{\gamma}
	\Big[ & 
	\alpha^{cd}\gamma^{AB}\gamma^{CD}
	\left(\mathcal{K}_{cAB}\mathcal{K}_{dCD} 
	- \mathcal{K}_{cAC}\mathcal{K}_{dBD}
	\right)
	- \frac{1}{4}\alpha^{ac}\alpha^{bd}\gamma_{CD}
		\mathcal{F}^C_{ab}\mathcal{F}^D_{cd}
	\Big] . 
\end{align}  

	We will now rewrite Eq.~\eqref{eq:S2odd} by integrating
over the two sphere. Firstly, we record the components of
$\mathcal{K}_{cAB}$ and $\mathcal{F}^C_{ac}$ subject to 
the perturbation Eq.~\eqref{RWAxial} 
\begin{align}
	\mathcal{K}_{cAB} 
	= & 
	\frac{\partial_cr}{r}\gamma_{AB} 
	- \frac{r^2}{2}\sum_{lm}h_c^{lm}
		\left(D_AB_B^{lm}+D_BB_A^{lm}\right) 
	, \\
	\mathcal{F}^C_{ab} 
	= & 
	\sum_{lm} (B^{lm})^C 
		\left(\partial_ah_b^{lm}-\partial_bh_a^{lm}\right) 
	+ \mathcal{O}\left((h_a^{lm})^2\right)
	,
\end{align}
	where $D_A$ is the covariant derivative on the two sphere (see 
Appendix \ref{SVTSpherical}).   
Using the properties of 
the axial vector spherical harmonics recorded
in Appendix~\ref{SVTSpherical}, and 
after several integrations by parts we obtain for the first two terms 
in Eq.~\eqref{eq:S2odd} as
\begin{align}
\label{Ksum}	
	& \int d^2\Omega \sqrt{\gamma}  
	\alpha^{cd}\gamma^{AB}\gamma^{CD}
	\left(\mathcal{K}_{cAB}\mathcal{K}_{dCD} 
	- \mathcal{K}_{cAC}\mathcal{K}_{dBD}
	\right)
	\nonumber \\ = &  
	2\alpha^{ab}\partial_ar\partial_br 
	- \frac{r^2}{2} 
	\sum_{l,m} l(l+1)\left[l(l+1)-2\right]\alpha^{ab}h^{lm}_ah^{lm}_b . 
\end{align}
	We drop the order zero term $\alpha^{ab}\partial_ar\partial_br$.
We next dimensionally reduce the `field strength' term 
(the one that depends on $(F^D_{ab})^2$) and obtain
\begin{equation}
\label{Fsum}
	\int d^2\Omega \sqrt{\gamma} 
	\gamma_{AB}\alpha^{ab}\alpha^{cd}F^A_{ac}F^B_{bd}
	 = r^4 \sum_{l,m} l(l+1) \alpha^{ab}\alpha^{cd}
	\mathcal{F}^{lm}_{ac}\mathcal{F}^{lm}_{bd}, 
\end{equation}
	where we have defined
\begin{equation}
\label{Flmab}
	\mathcal{F}^{lm}_{ab} \equiv \partial_ah^{lm}_b - \partial_bh^{lm}_a
	= {}^{\perp}\nabla_ah_b^{lm} - {}^{\perp}\nabla_bh^{lm}_a \,.
\end{equation}
	We can remove the factor of $r^4$ from Eq.~\eqref{Fsum} 
(multiplied by $\sqrt{-\alpha}$ in Eq.~\eqref{eq:S2odd}) by performing the 
following conformal transformation:
\begin{equation}
\label{conformalAxial}
	\hat{\alpha}_{ab} = \frac{1}{r^4}\alpha_{ab} .
\end{equation}
	Using Eqs.~\eqref{Ksum}--\eqref{conformalAxial}, we see that the 
dimensionally reduced Einstein-Hilbert action for axial perturbations 
about a spherically symmetric vacuum background is
\begin{equation}
\label{Saxial}
	S^{(2)}_{axial} = \sum_{l,m} l (l + 1) \int d^2u 
	\sqrt{-\hat{\alpha}^{lm}}
	\left[
	-\frac{1}{4}\hat{\alpha}^{ac}
	\hat{\alpha}^{bd}\mathcal{F}_{ab}^{lm}\mathcal{F}_{cd}^{lm} - 
	\frac{1}{2}M_{lo}^2(r)\hat{\alpha}^{ab}h_a^{lm}h_b^{lm}
	\right] ,
\end{equation}
	where we have defined an effective mass $M_{lo}(r)$ to be 
\begin{equation}
\label{Maxial}
	M_{lo}^2(r) =\left(l^2+l-2\right) r^2 .
\end{equation}

	The action in Eq.~\eqref{Saxial} is the central result of this section. 
We again note that up until this point the only condition we have placed on 
the two metric $\alpha_{ab}$ is that it has no functional dependence 
on the angular variables $\{\theta^A\}$. We conclude that the action 
(Eq.~\eqref{Saxial}) describes the linear metric axial perturbations of the 
Einstein-Hilbert action in a spherically symmetric vacuum background. 

We now derive the first order equations of motion by varying 
Eq.~\eqref{Saxial} with respect to $h_a^{lm}$:
\begin{equation}
\label{RW1}
	0 = {}^{\perp}\hat{\Box} h_a^{lm} 
	- \hat{\alpha}^{bc}{}^{\perp}\hat{\nabla}_b{}^{\perp}\hat{\nabla}_ah^{lm}_c 
- M_{lo}^2(r)h_a^{lm} .
\end{equation} 
	Here ${}^{\perp}\hat{\Box}\equiv\alpha^{ab}\nabla_a\nabla_b$ 
and ${}^{\perp}\hat{\nabla}_a$ are the derivative operators compatible with 
the background metric
constructed from $\hat\alpha_{ab}$ instead of $\alpha_{ab}$. 
Taking the 
divergence of Eq.~\eqref{RW1}, we obtain a constraint on the vector $h^{lm}_a$ 
as
\begin{equation}
\label{RW2}
	0 = {}^{\perp}\hat{\nabla}_a
	\left[M_{lo}^2(r)\hat{\alpha}^{ab}h^{lm}_b\right] . 
\end{equation}
	Recall that we may relate the Regge-Wheeler variable $h_a^{lm}$
to the gauge invariant variable under the simple substitution
$h_a^{lm}\to\tilde{h}_a^{lm}$, so that to linear order
in perturbation theory Eqs.~\eqref{Saxial},
\eqref{RW1}, and~\eqref{RW2}
under this relabeling become gauge invariant expressions. 

\subsubsection{\label{sec:masterAxialEOM}Master axial equation}
	For completeness, we demonstrate that we can 
rewrite Eqs.~\eqref{RW1},~\eqref{RW2} as a single
master equation (see, for example 
\cite{PhysRevD.19.2268,Sarbach:2001qq,SarbachThesis,
Martel:2005ir,MartelThesis}).
Firstly, we rewrite our equation of motion in the metric $\alpha_{ab}$. 
Note that as
$\sqrt{\hat{\alpha}}\hat{\alpha}^{ab}=\sqrt{\alpha}\alpha^{ab}$ and
$M^2_{lo}(r)=(l+1)(l-2) r^2$ where $(l+1)(l-2)$ is a constant, we see
that Eq.~\eqref{RW2} is equivalent to 
\begin{align}
\label{RW2_2}
	0 = {}^{\perp}\nabla_a\left(r^2\alpha^{ab}h^{lm}_b\right) .
\end{align} 
	We conclude that we can rewrite $h_b^{lm}$ in terms of the master 
variable for the odd parity perturbation $\Psi^{lm}_{\mathrm{odd}}$ as
\begin{align}
\label{h_vec_scalarVal}
	h_a^{lm} = 
	\frac{1}{r^2}\epsilon_{ab}
	{}^{\perp}\nabla^b\left(r\Psi^{lm}_{\mathrm{odd}}\right) ,
\end{align}
	where $\epsilon_{ab}$ is the Levi-Civita tensor\footnote{
We note that the Levi-Cevita tensor $\epsilon^{ab}$
is related to the Levi-Cevita symbol
$\tilde{\epsilon}^{ab}$
by $\epsilon^{ab}=\frac{-1}{\sqrt{-g}}\tilde{\epsilon}^{ab}$ for a Lorentzian
spacetime, so that $\epsilon_{ab}\epsilon^{bc}=+\delta_a{}^c$.} 
for the Lorentzian
metric $\alpha_{ab}$. Next, we rewrite Eq~\eqref{RW1} as
\begin{align}
\label{RW1_2}
	0 =
	{}^{\perp}\nabla^b
	\left[
	r^4 \left(
	{}^{\perp}\nabla_bh^{lm}_a - {}^{\perp}\nabla_ah^{lm}_b
	\right)
	\right]
	- (l-1)(l+2) r^2 h^{lm}_a 
	.
\end{align}
	In a two dimensional manifold we have the identity 
\begin{equation}
\label{2DAntiSymIdentity}
	2 \nabla^b\nabla_{[b}v_{a]}
	=
	\epsilon_{ab}\epsilon^{cd}\nabla^b\nabla_{[c}v_{d]}
	.
\end{equation} 
	 We use Eq.~\eqref{2DAntiSymIdentity}, 
along with Eq.~\eqref{h_vec_scalarVal} to
rewrite Eq.~\eqref{RW1_2} as
\begin{equation}
	0 =
	\epsilon_{ab}{}^{\perp}\nabla^b
	\left\{
	r^4\epsilon^{cd}{}^{\perp}\nabla_{[c}
	\left[
	\frac{1}{r^2}\epsilon_{d]p}
	{}^{\perp}\nabla^p\left(r\Psi^{lm}_{\mathrm{odd}}\right)
	\right]
	- (l-1)(l+2) r \Psi^{lm}_{\mathrm{odd}}
	\right\}
	.
\end{equation}	
	We integrate this equation and choose the integration constant
to be equal to zero. Expanding out our expression and using the background
equations of motion we obtain 
\begin{align}
\label{RW4}
	0 = 
	\left[
	{}^{\perp}\Box - \frac{l(l+1)}{r^2}+\frac{3}{2}{}^{\perp}\mathcal{R} 
	\right]
	\Psi^{lm}_{\mathrm{odd}}
	.
\end{align}
	We note that the master equation, Eq.~\eqref{RW4} only holds in a 
vacuum spacetime, for which we have the Schwarzschild background. 
For the background we can write  
${}^{\perp}\mathcal{R}=4M/r^3$, and we recover the Regge-Wheeler equation
\cite{PhysRev.108.1063} for axial perturbations.   
	
	We conclude that the variation of the dimensionally reduced action,
Eq.~\eqref{Saxial}, with respect to $h_a^{lm}$ gives us the
correct equations of motion for linear metric axial perturbations about 
a spherically symmetric vacuum spacetime. From these equations of motion
we are able to derive a covariant and gauge-invariant master equation
of motion for a scalar axial perturbation variable, as is done in, 
for example,
\cite{PhysRevD.19.2268,Sarbach:2001qq,SarbachThesis,
Martel:2005ir,MartelThesis}.

\subsection{Polar perturbations}
\label{PolarPert}
\subsubsection{Polar action}
Next, let us look at polar perturbations. In the Regge-Wheeler gauge, there 
are four nonzero polar perturbations: $\{ h_{ab}^{lm}, k^{lm}\}$. 
We begin by defining the following quantity 
\begin{equation}
\label{PhiDef}
	\Phi^2 = r^2e^{2k} ,
\end{equation}
	where (see Eq.~\eqref{gamABSphHarm})
\begin{equation}
	e^{2k} \equiv \sum_{l,m} e^{2k^{lm}} Y^{lm} .
\end{equation}
	We next define 
\begin{equation}
\label{phiDef}
	\phi^{lm} \equiv r \; \mathrm{exp}\left( k^{lm}\right) , 
\end{equation}
	 so that 	
\begin{equation}
\label{PhiSum}
	\Phi^2 = \sum_{l,m} \left(\phi^{lm}\right)^2 Y^{lm} . 
\end{equation}
	Using Eq.~\eqref{PhiSum}, we can write the line element for a spherically 
symmetric spacetime with polar perturbations as
\begin{equation}
\label{polarMetPert}
	ds^2 = \left({}^{(0)}\alpha_{ab} + \delta\alpha_{ab}\right)du^adu^b + 
\Phi^2d^2\Omega \equiv \alpha_{ab}du^adu^b + \Phi^2d^2\Omega . 
\end{equation} 
	With the metric in Eq.~\eqref{polarMetPert} at hand, we now derive the 
dimensionally reduced Einstein-Hilbert action. First we will look at the 
terms which depend on $\mathcal{K}_{cAB}$ and $\mathcal{A}_C^{ab}$, which in 
the metric Eq.~\eqref{polarMetPert} evaluate to be
\begin{align}
	\mathcal{K}_{cAB} = & \Omega_{AB}
		\sum_{l,m}Y^{lm}\phi^{lm}\partial_c\phi^{lm} 
	, \\
	\mathcal{A}^{ab}_C = & \frac{1}{2}\sum_{l,m}(h^{lm})^{ab}\partial_CY^{lm}
	.
\end{align}
	Integrating over the two sphere we obtain
\begin{align}
\label{KReduce}
	\int d^2ud^2\Omega\sqrt{-\alpha}\sqrt{\gamma}
	\left(
	\mathcal{K}_d\mathcal{K}^d - \mathcal{K}_{dAB}\mathcal{K}^{dAB}
	\right) 
= & \sum_{l,m} 2\int d^2u\sqrt{-\alpha^{lm}}\left(\partial_d\phi^{lm}\right)^2 
, \\
\label{AReduce}
	\int d^2ud^2\Omega\sqrt{-\alpha}\sqrt{\gamma}
	\left(
	\mathcal{A}_D\mathcal{A}^D - \mathcal{A}_{Dab}\mathcal{A}^{Dab}
	\right) 
= & \sum_{l,m} \frac{l(l+1)}{4} \int d^2u\sqrt{-\alpha^{lm}} 
	\left[
	\left(h^{lm}\right)^2 - \left( h_{ab}^{lm}\right)^2 
	\right] , 
\end{align} 
	where $h^{lm}\equiv{}^{(0)}\alpha^{ab} h_{ab}^{lm}$. 
Note that the dimensionally reduced action for the $\mathcal{A}^D_{ab}$ 
terms in Eq.~\eqref{AReduce} is the Fierz-Pauli graviton mass
~\cite{Fierz:1939ix}. We next compute ${}^{\parallel}\mathcal{R}$; firstly
we compute 
\begin{align}
	\Phi^2{}^{\parallel}\mathcal{R} 
	= 
	2 
	- 2 \Omega^{AB}D_AD_B\mathrm{ln}\Phi ,  
\end{align} 
	where $D_A$ is the covariant derivative for the round metric $\Omega_{AB}$
(see Appendix \ref{SVTSpherical}). Expanding $\Phi$ in terms of spherical
harmonics and to second order in the perturbations $k^{lm}$,  
$\Phi = r e^{k} 
	= r\sum_{lm}Y^{lm}e^{k^{lm}}
	=r\sum_{lm}Y^{lm}(1+k^{lm}+\frac{1}{2}(k^{lm})^2)$, 
and integrating over the two sphere we obtain
\begin{equation}
\label{RparaReduce}
	\int d^2ud^2\Omega \sqrt{-\alpha}\sqrt{\gamma}\,\, 
	{}^{\parallel}\mathcal{R} = 
	\sum_{l,m}\int d^2u\sqrt{-\alpha^{lm}}
	\left[2 + 2 l(l+1)  k^{lm} + \mathcal{O}\left((k^{lm})^3\right) \right] . 
\end{equation} 
	
	As all the terms in $h^{lm}_{ab}$ are scalars under the $SO(3)$ 
group action, 
we can straightforwardly dimensionally reduce ${}^{\perp}\mathcal{R}$:
\begin{equation}
\label{RperpReduce}
	\int d^2ud^2\Omega\sqrt{-\alpha}\sqrt{\gamma}\,\, 
	{}^{\perp}\mathcal{R} 
	= \sum_{l,m} 
	\int d^2u \sqrt{-\alpha^{lm}} \; (\phi^{lm})^2\,\, 
	{}^{\perp}\mathcal{R}^{lm} . 
\end{equation}
	In Eq.~\eqref{RperpReduce} we have not expanded out 
${}^{\perp}\mathcal{R}$ into a
 background piece and pieces linear and quadratic in the perturbation 
$ h_{ab}^{lm}$. 
Combining Eqs.~\eqref{KReduce}--\eqref{RperpReduce}, we obtain 
the dimensionally reduced action for linear polar perturbations of a 
spherically symmetric vacuum background in Regge-Wheeler gauge given by
\begin{align}
\label{Spolar}
	S^{(2)}_{polar} = \sum_{l,m}\int d^2u \sqrt{-\alpha^{lm}}
	\Bigg\{ &  
	\frac{(\phi^{lm})^2}{2}\,\, {}^{\perp}\mathcal{R}^{lm}  
	+ \left(\partial_d\phi^{lm}\right)^2 + 1 \nonumber \\
	& - \frac{l(l+1)}{8}\left[ ( h_{ab}^{lm})^2 
	- (h^{lm})^2\right] 
	+ l(l+1) k^{lm} 
	\Bigg\} . 
\end{align}
	Equation~\eqref{Spolar} is the action for a (1+1)-dimensional dilaton 
massive gravity model (see~\cite{deRham:2010kj} for another example of such 
a model, but without a dilaton field). Note that by setting 
$ h_{ab}^{lm}=0$ and $ k^{lm}=0$, the action in~\eqref{Spolar} 
reduces to the standard dimensionally reduced gravity action for a 
spherically symmetric vacuum background, Eq.~\eqref{EHAction_unperturbed}
(see for example Appendix B of
~\cite{frolov2012black}). For notational purposes, it is simpler to combine 
the linear and quadratic perturbations into the same action, and in 
Eq.~\eqref{Spolar} we have \emph{not} expanded out ${}^{\perp}\mathcal{R}^{lm}$ 
or $\phi^{lm}$ into a background plus linear perturbation.  
 
	We next derive the equations of motion that describe the dynamics of 
polar metric perturbations about a spherically symmetric vacuum background. 
In the 
equations of motion one can disentangle the background and perturbation 
degrees of freedom more easily than in the action. 
	Varying Eq.~\eqref{Spolar} by $ k^{lm}$, we have 
\begin{align}
\label{deltaphipolar}
	0 = (\phi^{lm})^2 \,\, 
	{}^{\perp}\mathcal{R}^{lm} - 2\phi^{lm}{}^{\perp}\Box\phi^{lm} 
	+ \frac{1}{2}l(l+1) h^{lm} . 
\end{align}
	Here we have defined 
$h^{ab} \equiv {}^{(0)}\alpha^{ac}\,{}^{(0)}\alpha^{bd}h_{cd}$, and
$h\equiv{}^{(0)}\alpha^{ab}h_{ab}$.
	The derivative operators ${}^{\perp}\nabla_a$ are 
treated as covariant derivative operators compatible
with the metric $\alpha_{ab}={}^{(0)}\alpha_{ab}+\delta\alpha_{ab}$.
(see Appendix \ref{factorizableSpacetime} for a discussion of 
the $m+n$ formalism and factorizable spacetimes). 
Three more independent equations of motion are derived by varying 
Eq.~\eqref{Spolar} by $(\alpha^{lm})^{ab}$, 
\begin{align}
\label{deltaalphapolar}
	0 = & 
	\Big[
	\frac{1}{2}(\partial\phi^{lm})^2 
	+ \phi^{lm}{}^{\perp}\Box\phi^{lm} 
	- \frac{1}{2}l(l+1) k^{lm} - \frac{1}{2}
	\Big]\alpha_{ab} 
	\nonumber \\ 
& -\phi^{lm}{}^{\perp}\nabla_a{}^{\perp}\nabla_b\phi^{lm} 
	 + \frac{l(l+1)}{4}\left( h_{ab}^{lm}-\alpha_{ab}h^{lm}\right) . 
\end{align}
 	We have not fully expanded out the metric, covariant derivatives, 
and $\phi\equiv re^k$ in this expression.
The right hand side of Eqs.~\eqref{deltaphipolar} and
~\eqref{deltaalphapolar} can be related to certain combinations of components 
of the full four dimensional Einstein tensor $G_{\mu\nu}$. Namely, 
Eq.~\eqref{deltaphipolar} corresponds to 
$-(r\,Y^{lm})^{-1}\left(G_{\theta\theta} + 
G_{\phi\phi}/\mathrm{sin}^{2}\theta\right)$, while 
Eq.~\eqref{deltaalphapolar} corresponds to $r^2(2 Y^{lm})^{-1}G_{ab}$. 
	We recall that the Regge-Wheeler variables $h_{ab}^{lm}$ and $k^{lm}$
can be related to the gauge invariant variables $\tilde{h}_{ab}^{lm}$
and $\tilde{k}^{lm}$ with the simple substitution 
$h_{ab}^{lm}\to\tilde{h}_{ab}^{lm}$ and $k^{lm}\to\tilde{k}^{lm}$, so
that to linear order in perturbation theory
Eqs.~\eqref{Spolar},~\eqref{deltaalphapolar}, and~\eqref{harmonicLike} 
under this relabeling are gauge invariant expressions.

\subsubsection{\label{sec:masterPolarEOM}Master polar equation}

	For completeness we demonstrate that we can rewrite
Eqs.~\eqref{deltaphipolar},~\eqref{deltaalphapolar}, and 
Eq.~\eqref{harmonicLike} as a single master equation
(see, for example \cite{Sarbach:2001qq,SarbachThesis,
Martel:2005ir,MartelThesis}). We set
\begin{align}
	r^a \equiv {}^{\perp}\nabla^ar ,
\end{align}
	where the $a$ index is raised/lowered with $\alpha^{ab}/\alpha_{ab}$, 
respectively. 

	We can take a divergence of Eq.~\eqref{deltaalphapolar}, and use 
Eq.~\eqref{deltaphipolar} to obtain the conditions
\begin{equation}
\label{harmonicLike}
	0 = 2{}^{\perp}\nabla_a k^{lm} 
		-{}^{\perp}\nabla^b h^{lm}_{ab} + {}^{\perp}\nabla_ah^{lm} 
		- \frac{{}^{\perp}\nabla_ar}{r}h^{lm} 
	.
\end{equation} 
	For higher dimensional massive gravity in flat space 
(for example, in $1+(d-1)$ dimensions), 
one can show that the addition of the Fierz-Pauli mass term to the 
Einstein-Hilbert action implies that the metric perturbation 
$\delta g_{\mu\nu}$ obeys a similar looking relation
~\cite{Hinterbichler:2011tt}, namely 
$\nabla^{\mu}\delta g_{\mu\nu} 
- g^{\alpha\beta}\nabla_{\nu}\delta g_{\alpha\beta}=0$. 

	We next expand out
Eq.~\eqref{deltaphipolar}. Using the background equations of
motion, Eqs~\eqref{deltarbkgrdEOM} and~\eqref{deltaalphabkgrdEOM}, along 
with Eq.~\eqref{harmonicLike} and  
the fact that in two dimensions $R_{ab}=\frac{1}{2}\alpha_{ab}R$, we see
that Eq.~\eqref{deltaphipolar} reduces to
\begin{align}
\label{deltakexpand}
	0 = & \frac{1}{2}\left[-2+l(1+1)\right] h^{lm} , 
\end{align}
	for $l>1$, we conclude that the metric perturbation is traceless. Lower
$l$ values require special treatment 
(e.g. \cite{Martel:2005ir,Sarbach:2001qq}); we do not consider $l=0,1$
in this article.  

	We expand out Eq.~\eqref{deltaalphapolar} to first order in metric
perturbations. Using $h=0$ and the background equations of motion, 
this reduces to 
\begin{align}
\label{deltaalphaexpand}
	0 = & 
	\left[
	- rr^c{}^{\perp}\nabla^dh^{lm}_{cd} - \frac{1}{2}r^cr^dh^{lm}_{cd}
	+ r^2{}^{\perp}\Box k^{lm} + 3 rr^c{}^{\perp}\nabla_ck 
	- \frac{1}{2}(l-1)(l+2)k^{lm}
	\right]
	\alpha_{ab}
	 \nonumber \\
	& 
	+ \frac{1}{2}rr^c
	\left(
	2{}^{\perp}\nabla_{(a}h^{lm}_{b)c}-{}^{\perp}\nabla_ch^{lm}_{ab}
	\right)	
	+ \frac{1}{4}\left[{}^{\perp}\mathcal{R}+l(l+1)\right]h^{lm}_{ab}
	- r^2{}^{\perp}\nabla_a{}^{\perp}\nabla_bk^{lm}
	- 2 r r_{(a}{}^{\perp}\nabla_{b)}k^{lm}
	.
\end{align}
	From Eqs.~\eqref{harmonicLike},~\eqref{deltakexpand}, 
and~\eqref{deltaalphaexpand}, we can construct the Zerilli-Moncrief
function, which is a covariant and gauge-invariant scalar which describes
the dynamics of the one independent polar degree of freedom. 
See, for example the discussions in 
\cite{Martel:2005ir,MartelThesis}
\footnote{Our Eq.~\eqref{deltaalphaexpand} is
equivalent to Eq.~(4.13) in \cite{Martel:2005ir} once we take into 
account the identity
\begin{equation}
	\nabla^c\nabla_{(a}p_{b)c} - \frac{1}{2}\Box p_{ab} 
	- \frac{1}{2}\alpha_{ab}\nabla^c\nabla^dp_{cd}
	= \frac{R}{2}p_{ab} ,
\end{equation}
	which holds for any traceless symmetric tensor $p_{ab}$ in a 
two dimensional manifold
\cite{MartelThesis,PhysRevD.19.2268}.}
. 
	The Zerilli-Moncrief function in our notation is
\begin{align}
	\Psi^{lm}_{\mathrm{even}}
	=
	\frac{2r}{l(l+1)}
	\left[
	2k^{lm} + \frac{2}{\Lambda}
	\left(r^ar^bh^{lm}_{ab} - 2rr^a{}^{\perp}\nabla_ak^{lm}\right)
	\right]
	,
\end{align}
	where we have defined \cite{Martel:2005ir} the function 
\begin{align}
	\Lambda = (l-1)(l+2)+\frac{3}{2}r^2{}^{\perp}\mathcal{R} .
\end{align}
	The Zerilli-Moncrief function obeys the Zerilli equation,
\begin{align}
	0 = \left({}^{\perp}\Box - V^{lm}_{\mathrm{even}}\right) 
	\Psi^{lm}_{\mathrm{even}}	
	,
\end{align}
	where
\begin{align}
	V^{lm}_{\mathrm{even}}
	=
	\frac{1}{\Lambda^2}
	\left[
	(l-1)^2(l+2)^2
	\left(\frac{(l-1)(l+2)+2}{r^2}+\frac{3}{2}{}^{\perp}\mathcal{R}\right)
	+ \frac{9}{4}r^2\left({}^{\perp}\mathcal{R}\right)^2
	\left((l-1)(l+2)+\frac{1}{2}r^2{}^{\perp}\mathcal{R}\right)
	\right]
	.
\end{align}
	Note that ${}^{\perp}\mathcal{R}=4M/r^3$ as the background
is a Schwarzschild black hole spacetime; substituting this value
in for ${}^{\perp}\mathcal{R}$ gives us a standard expression 
for the Zerilli potential. 
	We refer the reader to \cite{MartelThesis} for details on how 
to derive the Zerilli-Moncrief function and Zerilli equation from 
 Eqs.~\eqref{harmonicLike},~\eqref{deltakexpand}, 
and~\eqref{deltaalphaexpand}.

	We conclude that the variation of the dimensionally reduced action,
Eq.~\eqref{Spolar}, with respect to $k$ and $\alpha^{ab}$ gives us the
correct equations of motion for linear metric polar perturbations about 
a spherically symmetric vacuum spacetime, i.e. a Schwarzschild
black hole. From these equations of motion
we are able to derive a covariant and gauge-invariant master equation
of motion for a scalar axial perturbation variable, as is done in
\cite{Sarbach:2001qq,SarbachThesis,
Martel:2005ir,MartelThesis}.

\section{\label{sec:conclusion}Discussion and conclusion}

	In this work, we derived the action for linear perturbations about a 
spherically symmetric vacuum background in general relativity 
(Eqs.~\eqref{Saxial} and~\eqref{Spolar}) using a $2+2$ spacetime splitting. 
By dimensionally reducing the $2+2$ Einstein-Hilbert action to $(1+1)$ 
dimensions using the Regge-Wheeler gauge, we found that the axial 
perturbations are described by a massive vector field action 
(Eq.~\eqref{Saxial}), while the polar perturbations are described by a 
dilaton massive gravity action (Eq.~\eqref{Spolar}). Varying the 
actions Eqs.~\eqref{Saxial} and \eqref{Spolar}, we are able to rederive 
covariant
and gauge invariant master equations for the axial and polar degree of freedom,
respectively.  While in this article
we worked in a vacuum spacetime, with the addition of a cosmological constant
or matter source our results could be extended to study other backgrounds, 
such 
as the Schwarzschild (anti)-de Sitter
spacetime, or the Reissner-Nordstr\"{o}m spacetime.  

	To our knowledge, Eq.~\eqref{Spolar} is a novel (1+1)-dimensional 
massive gravity action (for another example of a two dimensional 
dilaton massive gravity model, see for example~\cite{deRham:2010kj}). 
The fact that we recover a massive gravity model from dimensionally 
reducing Einstein gravity may not come as a surprise: some four 
dimensional massive gravity models also arise from dimensionally reducing 
higher dimensional gravity theories
~\cite{Hinterbichler:2011tt,deRham:2014zqa}. 
One interesting feature of this model is that it describes
dynamics of linear gravitational waves about a Schwarzschild black hole. 
We note that since Schwarzschild black holes are classically 
stable to linear perturbations, the massive gravity theory as described 
by Eq.~\eqref{Spolar} is also classically linearly stable 
in that background. Two dimensional (dilaton) gravity has been used to study 
Hawking radiation and the quantum mechanics of black holes for 
`S-wave' scalar field perturbations
(see, for example,
~\cite{PhysRevD.14.870,1993stqg.conf..122H,Vagenas:2001sm}). The actions in 
Eqs.~\eqref{Saxial} and~\eqref{Spolar} could be useful in 
extending this program to investigating the quantum mechanics of gravitational
wave perturbations about Schwarzschild black holes; 
for example in constructing the path integral formulation of Hawking 
radiation for metric perturbations of a Schwarzschild black hole. 

	 The $m+n$ formalism is not limited to four dimensions and can 
be applied to a spacetime of arbitrary metric signature and arbitrary 
dimensionality. We caution that the $m+n$ formalism we present may be less 
useful 
in understanding the perturbations of spacetimes that cannot be foliated by 
subspaces that are maximally symmetric under the isometries of the full 
spacetime, i.e. spacetimes where one cannot write the background 
metric in the form of Eq.~\eqref{bkgrdSpherSym}. In these backgrounds 
the background frame vectors $n^{\alpha}_a$ do not form an 
involution
(e.g. $\beta^A_a\neq0$), the quantity $\alpha_{ab}$ is not the induced 
metric of a submanifold, and calculating and varying quantities 
such as ${}^{\perp}\mathcal{R}_{abcd}$ become much more cumbersome. 
	In particular, in the nonextremal Kerr spacetimes one cannot write 
the background 
metric in a form such that $\beta^{\alpha}_a=0$ on the background. 
Because of this fact, other formalisms such as the 
Newman-Penrose formalism~\cite{doi:10.1063/1.1724257} may ultimately remain 
more useful for understanding the dynamics and perturbations of backgrounds 
such as the nonextremal Kerr spacetime. 
	
\section*{Acknowledgments}

We thank 
Lasha Berezhiani,
Emanuele Berti, 
Eric Poisson, 
Frans Pretorius, and  
Teruaki Suyama
for fruitful and interesting discussions, reading through an 
earlier draft of this work, and for providing comments on that draft. 
	Additionally, we thank an anonymous referee, whose useful 
comments greatly helped us in improving our presentation 
of the $m+n$ formalism. We thank another referee for helpful comments
with regards to the master axial and polar equations of motion.
KY acknowledges support from JSPS Postdoctoral Fellowships for Research 
Abroad, NSF grant PHY-1305682 and the Simons Foundation.

\appendix

\section{\label{generalcoDimensionGeometry}Geometry of arbitrary 
codimension foliations}
	In this section we most closely follow the treatment of this subject 
by \cite{SouryaRayThesis2p2}; we review and extend their calculations 
here to set our
notation and to make this article more self-contained. 
	Assume that we have a $d$ dimensional manifold $M$
that has the topology
$\mathbb{R}^m\times\Sigma$. Furthermore, assume that $M$ can be foliated 
by an $n=d-m$ dimensional family of spacelike submanifolds
which we index with the label ${\bf t}\in\mathbb{R}^m$, 
$(\Sigma_{\bf t})_{{\bf t}\in\mathbb{R}^m}$. Greek indices 
will run from $0,...,d-1$. For any point $p\in M$,
the tangent space can split into
$T_p=T_p(\Sigma_{\bf t})\oplus T_p({}^{\perp}\Sigma_{\bf t})$, where
${}^{\perp}\Sigma_{\bf t}$ is called the transverse space to 
$\Sigma_{\bf t}$ and does
not generally integrate to form a submanifold.  
From now on we will drop the subscript ${\bf t}$ from 
$\Sigma_{\bf t}$ and ${}^{\perp}\Sigma_{\bf t}$; the use of the symbols
$\Sigma$ and ${}^{\perp}\Sigma$
will refer to a specific leaf of the foliation unless otherwise noted. 
We define
the tangent projection operator $h^{\mu}{}_{\nu}$ and the 
transverse projection
operator $l^{\mu}{}_{\nu} = \delta^{\mu}{}_{\nu} - h^{\mu}{}_{\nu}$
which project vectors $v^{\mu}\in T_p(M)$ to $T_p(\Sigma)$ and 
$T_p({}^{\perp}\Sigma)$, respectively. 
A tensor \emph{component} is called tangent if its contraction with
the transverse projector is zero; e.g. if $l_{\mu}{}^{\nu}P^{\mu\alpha}=0$ 
then we say the $\mu$ component of $P^{\mu\alpha}$ is tangent. Likewise
a component of a tensor is called transverse if its contraction
with the tangent projector is zero. A tensor is called 
tangent (transverse) if all of its components are tangent (transverse).
For example, consider a tensor 
$P^{\rho_1\cdots\mu_r}_{\sigma_1\cdots\sigma_s}$ at a point $p\in\Sigma$. 
This tensor is tangent to the leaf at this point if  
\begin{equation}
\label{eq:tangentTensor}
	h^{\mu_1}{}_{\rho_1}\cdots h^{\mu_r}{}_{\rho_r} h^{\sigma_1}{}_{\nu_1} \cdots
	h^{\sigma_s}{}_{\nu_s} P^{\rho_1\cdots \rho_r}_{\sigma_1\cdots \sigma_s}
	= P^{\mu_1\cdots \mu_r}_{\nu_1\cdots \nu_s}.
\end{equation} 
	and is transverse to the leaf at this point if 
\begin{equation}
\label{eq:tangentTensor}
	l^{\mu_1}{}_{\rho_1}\cdots l^{\mu_r}{}_{\rho_r} l^{\sigma_1}{}_{\nu_1} \cdots
	l^{\sigma_s}{}_{\nu_s} P^{\rho_1\cdots \rho_r}_{\sigma_1\cdots \sigma_s}
	= P^{\mu_1\cdots \mu_r}_{\nu_1\cdots \nu_s}.
\end{equation}

\subsection{Tangent/transverse derivatives and curvature tensors}

We next define tangent derivatives and tangent extrinsic curvature.
	We introduce a metric $g_{\mu\nu}$ and metric compatible covariant
derivative $\nabla_{\mu}$ on $M$. For tangent tensors 
$P^{\mu_1\cdots\mu_r}_{\nu_1\cdots\nu_s}\in T_p(\Sigma)^{\otimes r}\otimes 
T_p^*(\Sigma)^{\otimes s}$,  
the tangent derivative operator ${}^{\parallel}\nabla_{\mu}$ is defined as
the projection of the covariant derivative $\nabla_{\mu}$ by 
$h^{\mu}{}_{\nu}$
\begin{equation}
\label{eq:parallelDer}
	{}^{\parallel}\nabla_{\alpha}P^{\mu_1\cdots\mu_r}_{\nu_1\cdots\nu_s} 
	\equiv 
	h_{\alpha}{}^{\beta}
	h_{\rho_1}{}^{\mu_1}\cdots h_{\rho_r}{}^{\mu_r} 
	h_{\nu_1}{}^{\sigma_1}\cdots h_{\nu_s}{}^{\sigma_s}
	\nabla_{\beta}
	P^{\rho_1\cdots\rho_r}_{\sigma_1\cdots\sigma_s} .
\end{equation}
	The tangent extrinsic curvature $K^{\alpha}{}_{\mu\nu}$ can be defined as
follows. Consider $v^{\mu}\in T_p(\Sigma)$, then
\begin{equation}
\label{eq:extrinsicK_1}
	h_{\mu}{}^{\rho}\nabla_{\rho}v^{\nu} 
	\equiv {}^{\parallel}\nabla_{\mu}v^{\nu}
	- K^{\nu}{}_{\lambda\mu}v^{\lambda},	
\end{equation} 
	in other words we have
\begin{equation}
\label{eq:extrinsicK_2}
	K^{\nu}{}_{\lambda\mu} \equiv 
	h_{\lambda}{}^{\sigma}h_{\mu}{}^{\rho}
	\nabla_{\rho}l_{\sigma}{}^{\nu}. 
\end{equation}
	The tangent extrinsic curvature is also known as the second fundamental 
form. Following similar terminology to that of Carter 
\cite{1992JGP.....8...53C},
we write 
$K^{\lambda}\equiv K^{\lambda}{}_{\alpha}{}^{\alpha}$, which
we call the tangent curvature vector.
As $\Sigma$ is a submanifold, $K^{\lambda}{}_{\mu\nu}$ is
symmetric under $\mu\leftrightarrow\nu$; Carter
\cite{1992JGP.....8...53C} refers to this property as 
the generalized Weingarten-Frobenius identity .
From the definition in Eq.~\eqref{eq:extrinsicK_2} we see that
\begin{equation}
\label{eq:extrinsicK_properties}
	h_{\lambda}{}^{\sigma}K^{\lambda}{}_{\alpha\beta} = 
	l_{\alpha}{}^{\mu}K^{\lambda}{}_{\mu\beta} = 
	l_{\beta}{}^{\mu}K^{\lambda}{}_{\alpha\mu} = 0 \,. 
\end{equation} 

	The transverse derivative operator ${}^{\perp}\nabla_{\mu}$ and 
the transverse extrinsic curvature are defined in a similar manner to
what is done for ${}^{\parallel}\nabla_{\mu}$.
Consider a transverse tensor 
$P^{\mu_1\cdots\mu_r}_{\nu_1\cdots\nu_s}\in T_p({}^{\perp}\Sigma)^{\otimes r}
\otimes T_p^*({}^{\perp}\Sigma)^{\otimes s}$,  
then
\begin{equation}
\label{eq:perpDer}
	{}^{\perp}\nabla_{\alpha}P^{\mu_1\cdots\mu_r}_{\nu_1\cdots\nu_s} 
	\equiv 
	l_{\alpha}{}^{\beta}
	l_{\rho_1}{}^{\mu_1}\cdots l_{\rho_r}{}^{\mu_r} 
	l_{\nu_1}{}^{\sigma_1}\cdots l_{\nu_s}{}^{\sigma_s}
	\nabla_{\beta}
	P^{\rho_1\cdots\rho_r}_{\sigma_1\cdots\sigma_s} .
\end{equation}
	The transverse extrinsic curvature $A^{\alpha}{}_{\mu\nu}$ is 
defined as 
follows. Consider $v^{\mu}\in T_p({}^{\perp}\Sigma)$, then 
\begin{equation}
\label{eq:extrinsicA_1}
	l_{\mu}{}^{\rho}\nabla_{\rho}v^{\nu} 
	\equiv {}^{\perp}\nabla_{\mu}v^{\nu}
	- A^{\nu}{}_{\lambda\mu}v^{\lambda},	
\end{equation} 
	in other words we have
\begin{equation}
\label{eq:extrinsicA_2}
	A^{\nu}{}_{\lambda\mu} \equiv 
	l_{\lambda}{}^{\sigma}l_{\mu}{}^{\rho}
	\nabla_{\rho}h_{\sigma}{}^{\nu}. 
\end{equation}
	We write 
$A^{\lambda}\equiv A^{\lambda}{}_{\alpha}{}^{\alpha}$, which we call 
the transverse curvature vector.
From the definition in Eq.~\eqref{eq:extrinsicA_2} we see that
\begin{equation}
\label{eq:extrinsicA_properties}
	l_{\lambda}{}^{\sigma}A^{\lambda}{}_{\alpha\beta} = 
	h_{\alpha}{}^{\mu}A^{\lambda}{}_{\mu\beta} =   
	h_{\beta}{}^{\mu}A^{\lambda}{}_{\alpha\mu} = 0\,.  
\end{equation}	
	As the transverse space ${}^{\perp}\Sigma$ does not generally 
integrate to form a submanifold,  
the transverse extrinsic curvature $A^{\alpha}{}_{\mu\nu}$ is 
generally
not symmetric in $\mu\leftrightarrow\nu$. This is reflected by the fact that
the action of two 
transverse derivatives ${}^{\perp}\nabla_{\mu}$ on a scalar function $f$ 
generally do not commute. We define the transverse torsion 
tensor $F^{\lambda}{}_{\alpha\beta}$, where
\begin{align}
\label{eq:Torsion}
	F^{\lambda}{}_{\alpha\beta}\nabla_{\lambda}f \equiv & 
	-2{}^{\perp}\nabla_{[\alpha}{}^{\perp}\nabla_{\beta]}f  \\
	= & -2 A^{\lambda}{}_{[\alpha\beta]}\nabla_{\lambda}f .\nonumber 
\end{align}
	We see that the transverse torsion tensor is the antisymmetric 
component of 
the transverse extrinsic curvature $A^{\lambda}{}_{\mu\nu}$.
The transverse torsion tensor $F^{\lambda}{}_{\alpha\beta}$ is also 
known as the twist connection. 

	We now define the curvature tensors for the derivative operators
${}^{\parallel}\nabla_{\mu}$ and ${}^{\perp}\nabla_{\mu}$. Consider
a form $v_{\mu}\in T^*_p(\Sigma)$, we then define
\begin{align}
\label{eq:parallelR}
	{}^{\parallel}R_{\alpha\beta\gamma}{}^{\delta}v_{\delta} 
	\equiv & 
	2{}^{\parallel}\nabla_{[\alpha}{}^{\parallel}\nabla_{\beta]}
	v_{\gamma}  \\
	= & 2 h_{[\alpha}{}^{\mu}h_{\beta]}{}^{\nu}h_{\gamma}{}^{\lambda}
	\nabla_{\mu}\left(h_{\nu}{}^{\alpha}h_{\lambda}{}^{\rho}
		\nabla_{\alpha}v_{\rho}\right) . \nonumber 
\end{align}
	The curvature tensor for the operator ${}^{\perp}\nabla_{\mu}$
is defined similarly, except that we need to take into account that it 
generally will have nonzero torsion. 
Consider a form $v_{\mu}\in T_p^*({}^{\perp}\Sigma)$, we 
then define 
\begin{align}
\label{eq:perpR}
	{}^{\perp}R_{\alpha\beta\gamma}{}^{\delta}v_{\delta} \equiv & 
	2{}^{\perp}\nabla_{[\alpha}{}^{\perp}\nabla_{\beta]}
	v_{\gamma} + F^{\lambda}{}_{\alpha\beta}l_{\gamma}{}^{\delta}
		\nabla_{\lambda}v_{\delta} \\
	= & 2 l_{[\alpha}{}^{\mu}l_{\beta]}{}^{\nu}l_{\gamma}{}^{\lambda}
	\nabla_{\mu}\left(l_{\nu}{}^{\alpha}l_{\lambda}{}^{\rho}
		\nabla_{\alpha}v_{\rho}\right) 
	- 2A^{\lambda}{}_{[\alpha\beta]}l_{\gamma}{}^{\delta}
		\nabla_{\lambda}v_{\delta} \,. \nonumber 
\end{align}
	Note that the derivative acting on $v_{\delta}$ contracted with the 
torsion tensor is \emph{not} ${}^{\perp}\nabla_{\mu}$ as 
$l_{\lambda}{}^{\rho}A^{\lambda}{}_{\alpha\beta}=0$. 

\subsection{Projections of the Riemann tensor\label{ProjectionsRiemann}}
	With the definitions in
Eqs.~\eqref{eq:extrinsicK_2},~\eqref{eq:extrinsicA_2},~\eqref{eq:parallelR},
and \eqref{eq:perpR}, we can rewrite the projections of the Riemann tensor
by $h_{\mu}{}^{\nu}$ and $l_{\mu}{}^{\nu}$ entirely in terms of the tensors
$K^{\alpha}{}_{\mu\nu}$, $A^{\alpha}{}_{\mu\nu}$,
${}^{\parallel}R_{\alpha\beta\gamma\delta}$,
and ${}^{\perp}R_{\alpha\beta\gamma\delta}$.  
	These are summarized below:
\begin{align}
\label{eq:hhhhR}
	h_{\alpha}{}^{\mu}h_{\beta}{}^{\nu}h_{\gamma}{}^{\lambda}
h_{\delta}{}^{\rho}R_{\mu\nu\lambda\rho}  
	= & {}^{\parallel}R_{\alpha\beta\gamma\delta} - 
K^{\lambda}{}_{\gamma\alpha}K_{\lambda\delta\beta} 
+ K^{\lambda}{}_{\gamma\beta}K_{\lambda\delta\alpha} ,\\ 
\label{eq:llllR}
	l_{\alpha}{}^{\mu}l_{\beta}{}^{\nu}l_{\gamma}{}^{\lambda}
l_{\delta}{}^{\rho}R_{\mu\nu\lambda\rho}  
	= & {}^{\perp}R_{\alpha\beta\gamma\delta} - 
A^{\lambda}{}_{\gamma\alpha}A_{\lambda\delta\beta} + 
A^{\lambda}{}_{\gamma\beta}A_{\lambda\delta\alpha} ,\\ 
\label{eq:hhllR}
	l_{\alpha}{}^{\mu}h_{\beta}{}^{\nu}l_{\gamma}{}^{\lambda}
h_{\delta}{}^{\rho}R_{\mu\nu\lambda\rho} 
	= &- l_{\alpha}{}^{\mu}
h_{\beta}{}^{\nu}l_{\gamma}{}^{\lambda}h_{\delta}{}^{\rho}\nabla_{\mu}
K_{\lambda\rho\nu} - l_{\alpha}{}^{\nu}h_{\beta}{}^{\mu}
l_{\gamma}{}^{\lambda}h_{\delta}{}^{\rho}\nabla_{\mu}
A_{\rho\lambda\nu} \\
	& - K_{\alpha}{}^{\lambda}{}_{\beta}
K_{\gamma\delta\lambda} - A_{\beta}{}^{\lambda}{}_{\alpha}
A_{\delta\gamma\lambda} , \nonumber \\
\label{eq:hhhlR}
	h_{\alpha}{}^{\mu}h_{\beta}{}^{\nu}h_{\gamma}{}^{\lambda}
l_{\delta}{}^{\rho}R_{\mu\nu\lambda\rho} 
	= & 2 h_{[\alpha}{}^{\mu}h_{\beta]}{}^{\nu}h_{\gamma}{}^{\lambda}
		\nabla_{\mu}K_{\delta\lambda\nu} , \\ 
\label{eq:lllhR}
	l_{\alpha}{}^{\mu}l_{\beta}{}^{\nu}l_{\gamma}{}^{\lambda}
h_{\delta}{}^{\rho}R_{\mu\nu\lambda\rho} 
	= & 2l_{[\alpha}{}^{\mu}l_{\beta]}{}^{\nu}l_{\gamma}{}^{\lambda}
		\nabla_{\mu}A_{\delta\lambda\nu}
	+ 2 A^{\sigma}{}_{[\alpha\beta]}K_{\gamma\delta\sigma} . 
\end{align}
	Eq.~\eqref{eq:hhhhR} is the $m+n$ generalization of the Gauss equation, 
Eq.~\eqref{eq:hhllR} is the $m+n$ generalization of the Ricci equation,
and Eq.~\eqref{eq:hhhlR} is the $m+n$ generalization of the Codazzi equation.
Eqs.~\eqref{eq:llllR} and~\eqref{eq:lllhR} are identically zero in
codimension one spacetime splittings.
We provide a derivation of Eqs.~\eqref{eq:llllR} and~\eqref{eq:hhllR} 
below; the
derivation of the other projections follow a similar procedure. Similar
expressions projections
of the Riemann tensor are presented
in Appendix A of \cite{SouryaRayThesis2p2}.  

\subsubsection{\label{DerivationHHHHR}Derivation of Eq.~\eqref{eq:llllR}}
	To show Eq.~\eqref{eq:llllR}, let us consider 
$v^{\mu}\in T_p({}^{\perp}\Sigma)$. We then have 
\begin{align}
	l_{\alpha}{}^{\mu}l_{\beta}{}^{\nu}l_{\gamma}{}^{\lambda}
	R_{\mu\nu\lambda\sigma}v^{\sigma}
	= &  
	2l_{[\alpha}{}^{\mu}l_{\beta]}{}^{\nu}l_{\gamma}{}^{\lambda}
	\nabla_{\mu}\nabla_{\nu}v_{\lambda}
	 \nonumber \\
	= &  
	2l_{[\alpha}{}^{\mu}l_{\beta]}{}^{\nu}l_{\gamma}{}^{\lambda}\nabla_{\mu}
	\left[
	\left(l_{\nu}{}^{\rho}+h_{\nu}{}^{\rho}\right)
	\left(l_{\lambda}{}^{\sigma}+h_{\lambda}{}^{\sigma}\right)
	\nabla_{\rho}v_{\sigma}
	\right] 
	\nonumber \\
	= & 
	2A^{\rho}{}_{[\beta\alpha]}l_{\gamma}{}^{\sigma}\nabla_{\rho}v_{\sigma}
	+ 2l_{[\alpha}{}^{\mu}l_{\beta]}^{\rho}l_{\gamma}{}^{\lambda}
	\nabla_{\mu}h_{\lambda}{}^{\sigma}\nabla_{\rho}v_{\sigma}
	+ 2 {}^{\perp}\nabla_{[\alpha}{}^{\perp}\nabla_{\beta]}v_{\gamma} 
	 \nonumber \\
	= & 
	2\,{}^{\perp}\nabla_{[\alpha}{}^{\perp}\nabla_{\beta]}v_{\gamma}	
	+ F^{\rho}{}_{\alpha\beta}l_{\gamma}{}^{\sigma}\nabla_{\rho}v_{\sigma}
	- 2 l_{[\alpha}{}^{\mu}l_{\beta]}{}^{\rho}h_{\lambda}{}^{\sigma}
	\left(\nabla_{\mu}l_{\gamma}{}^{\lambda}\right)
	\nabla_{\rho}\left(l_{\sigma}{}^{\kappa}v_{\kappa}\right) 
	 \nonumber \\
	= & 
	  {}^{\perp}R_{\alpha\beta\gamma\delta}v^{\delta}
	+ 2 l_{[\alpha}{}^{\mu}l_{\beta]}{}^{\rho} l_{\sigma\delta}
	  \left(\nabla_{\mu}l_{\gamma}{}^{\lambda}\right)
	  \left(\nabla_{\rho}h_{\lambda}{}^{\sigma}\right) v^{\delta} .
\end{align}
	Consider the last term: 
\begin{align}
	\left( 
	l_{\alpha}{}^{\mu}l_{\beta}{}^{\rho}-l_{\beta}{}^{\mu}l_{\alpha}^{\rho}
	\right)
	l_{\sigma\delta}
	\left(\nabla_{\mu}l_{\gamma}{}^{\lambda}\right)
	\left(\nabla_{\rho}h_{\lambda}{}^{\sigma}\right)v^{\delta}
	= & 
	\left[
	\left(
	h_{\kappa}{}^{\lambda}l_{\alpha}{}^{\mu}\nabla_{\mu}l_{\gamma}{}^{\kappa}
	\right)
	A_{\lambda\delta\beta}
	-
	\left(
	h_{\kappa}{}^{\lambda}l_{\beta}{}^{\mu}\nabla_{\mu}l_{\gamma}{}^{\kappa}
	\right)
	A_{\lambda\delta\alpha}
	\right]
	v^{\delta}
	\nonumber \\
	= & 
	\left(
	- A^{\lambda}{}_{\gamma\alpha}A_{\lambda\delta\beta}	
	+ A^{\lambda}{}_{\gamma\beta}A_{\lambda\delta\alpha}
	\right) v^{\delta} .
\end{align}
	We conclude that Eq.~\eqref{eq:llllR} holds, 
\begin{equation}
	l_{\alpha}{}^{\mu}l_{\beta}{}^{\nu}l_{\gamma}{}^{\lambda}
l_{\delta}{}^{\rho}R_{\mu\nu\lambda\rho}  
	= {}^{\perp}R_{\alpha\beta\gamma\delta} - 
A^{\lambda}{}_{\gamma\alpha}A_{\lambda\delta\beta} + 
A^{\lambda}{}_{\gamma\beta}A_{\lambda\delta\alpha} . 	
\end{equation}

\subsubsection{\label{DerivationHHLLR}Derivation of Eq.~\eqref{eq:hhllR}}
	To show Eq.~\eqref{eq:hhllR}, let us consider $v^{\mu}\in T_p(\Sigma)$. 
We compute 
\begin{align}
\label{eq:hhllR_Inter1}
	l_{\alpha}{}^{\mu}h_{\beta}{}^{\nu}l_{\gamma}{}^{\lambda}
	R_{\mu\nu\lambda\delta}v^{\delta} 
	= & 2 l_{\alpha}{}^{[\mu}h_{\beta}{}^{\nu]}l_{\gamma}{}^{\lambda}
	\nabla_{\mu}\nabla_{\nu}v_{\lambda} 
	 \nonumber \\
	= & 2 l_{\alpha}{}^{[\mu}h_{\beta}{}^{\nu]}l_{\gamma}{}^{\lambda}
	\Big(
	  (\nabla_{\mu}h_{\lambda}{}^{\rho})(\nabla_{\nu}v_{\rho})
	+ (\nabla_{\nu}h_{\lambda}{}^{\rho})(\nabla_{\mu}v_{\rho})
	\nonumber \\
	& + h_{\lambda}{}^{\rho}\nabla_{\mu}\nabla_{\nu}v_{\rho}
	+ v_{\rho}\nabla_{\mu}\nabla_{\nu}h_{\lambda}{}^{\rho}
	\Big) 
	 \nonumber \\
	= & 2l_{\alpha}{}^{[\mu}h_{\beta}{}^{\nu]}l_{\gamma}{}^{\lambda}
	\left(\nabla_{\mu}\nabla_{\nu}h_{\lambda}{}^{\rho}\right)v_{\rho}
	\nonumber \\
	= & 2 
	l_{\alpha}{}^{[\mu}h_{\beta}{}^{\nu]}l_{\gamma}{}^{\sigma}
	\left[
	\nabla_{\mu}
	\left(l_{\sigma}{}^{\lambda}\nabla_{\nu}h_{\lambda\delta}\right) 
	- \left(\nabla_{\mu}l_{\sigma}{}^{\lambda}\right)
	  \left(\nabla_{\nu}h_{\lambda\delta}\right)
	\right] v^{\delta} . 
\end{align}
	We next split this calculation into two different parts. 
We first look at 
\begin{align}
	2 l_{\alpha}{}^{[\mu}h_{\beta}{}^{\nu]}l_{\gamma}{}^{\sigma}
	\left[	
	\nabla_{\mu}
	\left(
	l_{\sigma}{}^{\lambda}\nabla_{\nu}h_{\lambda\delta}
	\right) \right]v^{\delta}
	= & 2 l_{\alpha}{}^{[\mu}h_{\beta}{}^{\nu]}l_{\gamma}{}^{\sigma}
	\left[
	\nabla_{\mu}
	\left(	
	\left(
	h_{\nu}{}^{\kappa}+l_{\nu}{}^{\kappa}
	\right)
	l_{\sigma}{}^{\lambda}\nabla_{\kappa}
	h_{\lambda\delta}
	\right)
	\right] v^{\delta}
	\nonumber \\	
	= & 
	2 l_{\alpha}{}^{[\mu}h_{\beta}{}^{\nu]}l_{\gamma}{}^{\sigma}
	\left[\nabla_{\mu}
	\left(
	A_{\delta\sigma\nu} - K_{\sigma\delta\nu}
	\right) \right] v^{\delta} ,
\end{align}
	where we have used 
$\delta_{\nu}{}^{\kappa}=l_{\nu}{}^{\kappa}+h_{\nu}{}^{\kappa}$.
	We further split this term into two more pieces
\begin{align}
	2l_{\alpha}{}^{[\mu}h_{\beta}{}^{\nu]}l_{\gamma}{}^{\sigma}
	\left(\nabla_{\mu}A_{\delta\sigma\nu}\right)v^{\delta}
	= & 
	\left[
	- l_{\alpha}{}^{\nu}h_{\beta}{}^{\mu}
	  l_{\gamma}{}^{\sigma}h_{\delta}{}^{\lambda}
	  \nabla_{\mu}A_{\lambda\sigma\nu}
	+ l_{\alpha}{}^{\mu}h_{\beta}{}^{\nu} 
	  \left(\nabla_{\mu}l_{\nu}{}^{\xi}\right)
	  A_{\delta\gamma\xi}
	\right] v^{\delta}	
	 \nonumber \\
	= & 
	\left(
		- l_{\alpha}{}^{\nu}h_{\beta}{}^{\mu}
		  l_{\gamma}{}^{\sigma}h_{\delta}{}^{\lambda}
		  \nabla_{\mu}A_{\lambda\sigma\nu}
		- A_{\delta\gamma\lambda}A_{\beta}{}^{\lambda}{}_{\alpha}
	\right) v^{\delta} . 
\end{align} 
	Similarly we have	
\begin{align}
	2l_{\alpha}{}^{[\mu}h_{\beta}{}^{\nu]}l_{\gamma}{}^{\sigma}
	\left(\nabla_{\mu}K_{\sigma\delta\nu}\right)v^{\delta}
	= 
	\left(
	  l_{\alpha}{}^{\mu}h_{\beta}{}^{\nu}
	  l_{\gamma}{}^{\sigma}h_{\delta}{}^{\lambda}
	  \nabla_{\mu}K_{\sigma\lambda\nu}
	+ K_{\alpha}{}^{\lambda}{}_{\beta}K_{\gamma\delta\lambda}
	\right) v^{\delta} .
\end{align}
	Finally, we look at the last term on the right hand side of 
Eq.~\eqref{eq:hhllR_Inter1}, 
\begin{align}
	2l_{\alpha}{}^{[\mu}h_{\beta}{}^{\nu]}l_{\gamma}{}^{\sigma}
	\left(\nabla_{\mu}l_{\sigma}{}^{\lambda}\right)
	\left(\nabla_{\nu}h_{\lambda\delta}\right) v^{\delta}
	= & 
	2l_{\alpha}{}^{[\mu}h_{\beta}{}^{\nu]}l_{\gamma}{}^{\sigma}
	\left(\nabla_{\mu}h_{\sigma}{}^{\lambda}\right)
	\left(\nabla_{\nu}l_{\lambda\delta}\right) v^{\delta}
	\nonumber \\
	= & 
	2l_{\alpha}{}^{[\mu}h_{\beta}{}^{\nu]}
	l_{\lambda\delta}
	\left(\nabla_{\mu}l_{\gamma}{}^{\sigma}\right)
	\left(\nabla_{\nu}h_{\sigma}{}^{\lambda}\right) v^{\delta}
	\nonumber \\
	= & 0 ,	
\end{align}
	which is zero as $v^{\mu}$ is a tangent vector.
With this final relation we can recover Eq.~\eqref{eq:hhllR},
\begin{align}
	l_{\alpha}{}^{\mu}h_{\beta}{}^{\nu}l_{\gamma}{}^{\lambda}
h_{\delta}{}^{\rho}R_{\mu\nu\lambda\rho} 
	= &- l_{\alpha}{}^{\mu}
h_{\beta}{}^{\nu}l_{\gamma}{}^{\lambda}h_{\delta}{}^{\rho}\nabla_{\mu}
K_{\lambda\rho\nu} - l_{\alpha}{}^{\nu}h_{\beta}{}^{\mu}
l_{\gamma}{}^{\lambda}h_{\delta}{}^{\rho}\nabla_{\mu}
A_{\rho\lambda\nu} \nonumber \\
	& - K_{\alpha}{}^{\lambda}{}_{\beta}
K_{\gamma\delta\lambda} - A_{\beta}{}^{\lambda}{}_{\alpha}
A_{\delta\gamma\lambda} .
\end{align}

\subsection{\label{ProjRicciScalar}Projected Ricci tensor
and projected Ricci scalar}

	Using Eqs.~\eqref{eq:hhhhR},~\eqref{eq:llllR}, and \eqref{eq:hhllR}, 
we can
rewrite the Ricci tensor in terms of 
${}^{\parallel}R_{\alpha\beta\gamma\delta}$, 
${}^{\perp}R_{\alpha\beta\gamma\delta}$, 
$A_{\gamma\alpha\beta}$, and 
$K_{\gamma\alpha\beta}$. Using the completeness relation 
$g_{\mu\nu}=l_{\mu\nu}+h_{\mu\nu}$, we have
\begin{align}
\label{eq:hhRicciTensor}
	h_{\alpha}{}^{\mu}h_{\beta}{}^{\lambda}R_{\mu\lambda}
	= & 
	h_{\alpha}{}^{\mu}h_{\beta}{}^{\lambda}
	\left(h^{\nu\rho} + l^{\nu\rho}\right)
	R_{\mu\nu\lambda\rho}
	\nonumber \\
	= & 
	-h_{\alpha}{}^{\mu}h_{\beta}{}^{\lambda}l^{\nu\rho}
		\nabla_{\nu}K_{\rho\lambda\mu}
	- h_{\alpha}{}^{\nu}h_{\beta}{}^{\lambda}l^{\mu\rho}
		\nabla_{\nu}A_{\lambda\rho\mu}
	\nonumber \\
	& -  K^{\lambda}{}_{\beta\alpha}K_{\lambda} 
	- A_{\alpha}{}^{\lambda\sigma}A_{\beta\sigma\lambda}
	+ {}^{\parallel}R_{\alpha\beta} 
	, \\
\label{eq:llRicciTensor}
	l_{\alpha}{}^{\mu}l_{\beta}{}^{\lambda}R_{\mu\lambda} 
	= &
	l_{\alpha}{}^{\mu}l_{\beta}{}^{\lambda}
	\left(l^{\nu\rho}+h^{\nu\rho}\right)R_{\mu\nu\lambda\rho} 
	 \nonumber \\
	= &
	- l_{\alpha}{}^{\mu}l_{\beta}{}^{\lambda}h^{\nu\rho}
		\nabla_{\nu}A_{\rho\lambda\mu}
	- l_{\alpha}{}^{\nu}l_{\beta}{}^{\lambda}h^{\mu\rho}
		\nabla_{\nu}K_{\lambda\rho\mu}
	\nonumber \\
	& - A^{\lambda}{}_{\beta\alpha}A_{\lambda} 
	- K_{\alpha}{}^{\lambda\sigma}K_{\beta\sigma\lambda} 
	+ {}^{\perp}R_{\alpha\beta}
	, \\
\label{eq:hlRicciTensor}
	h_{\alpha}{}^{\nu}l_{\beta}{}^{\rho}R_{\nu\rho}
	= & 
	h_{\alpha}{}^{\nu}l_{\beta}{}^{\rho}
	\left(h^{\mu\lambda} + l^{\mu\lambda}\right)
	R_{\mu\nu\lambda\rho}	
	 \nonumber \\
	= & 2 l_{[\sigma}{}^{\mu}l_{\beta]}{}^{\nu}l^{\sigma\lambda}
		\nabla_{\mu}A_{\alpha\lambda\nu}
	+ 2 h_{[\sigma}{}^{\mu}h_{\alpha]}{}^{\nu}h^{\sigma\lambda}
		\nabla_{\mu}K_{\beta\lambda\nu}
	+ 2 A^{\sigma}{}_{[\mu\beta]}K^{\mu}{}_{\alpha\sigma} 
	.
\end{align}
	Eqs.~\eqref{eq:hhRicciTensor}, \eqref{eq:llRicciTensor}, and 
\eqref{eq:hlRicciTensor} are the projected vacuum Einstein equations. 
In the context of a double null foliation in four dimensional
spacetime (see section \ref{nullFoliation}), some authors have pointed out 
that Eq.~\eqref{eq:hlRicciTensor}, with a suitable relabeling 
and interpretation of its 
variables resembles a Navier-Stokes equation
\cite{Gourgoulhon:2005ch,
1982mgm..conf..587D,PhysRevD.18.3598,DamourThesis} 
(see, e.g. \cite{Padmanabhan:2010rp} for a critique of this
interpretation).
 
	Calculating one further contraction gives us the 
projected Ricci scalar,
\begin{align}
\label{eq:projRicciScalar}
	R = & \left(
	h^{\alpha\gamma}h^{\beta\delta} + l^{\alpha\gamma}l^{\beta\delta}
	+ 2 h^{\alpha\gamma}l^{\beta\delta}
		\right) 
	R_{\alpha\beta\gamma\delta}  \nonumber \\
	= & {}^{\parallel}R 
	  + {}^{\perp}R 
	+ K_{\lambda}K^{\lambda}		
	- K_{\lambda\alpha\beta}K^{\lambda\beta\alpha}
	+ A_{\lambda}A^{\lambda}		
	- A_{\lambda\alpha\beta}A^{\lambda\beta\alpha}
	  - 2 \nabla_{\lambda}\left(K^{\lambda} 
			+ A^{\lambda}\right) . 
\end{align} 
	One can similarly apply the Riemann projection formulas to rewrite
scalar polynomials in the Riemann curvature, such as
$R_{\alpha\beta\gamma\delta}R^{\alpha\beta\gamma\delta}$ in terms of the
quantities  
${}^{\parallel}R_{\alpha\beta\gamma\delta}$, 
${}^{\perp}R_{\alpha\beta\gamma\delta}$, 
$A_{\gamma\alpha\beta}$, and 
$K_{\gamma\alpha\beta}$. 

\subsection{\label{codimensionOne} Codimension one foliations} 
	Let us now consider a special case with codimension one foliations.
	For a codimension one surface, we can write 
$l^{\mu\nu}=\epsilon n^{\mu}n^{\nu}$. We choose $n^{\mu}$ to be normalized
to $\epsilon\equiv\pm1$ depending on whether $n^{\mu}$ is space- or time-like. 
The completeness relation for the projection operators
then reads
\begin{equation}
\label{eq:coDimOne_Completeness}
	g_{\mu\nu} = h_{\mu\nu} + \epsilon n_{\mu}n_{\nu} .
\end{equation} 
In this case, we see that 
\begin{align}
\label{eq:coDimOne_K}
	K^{\lambda}{}_{\alpha\beta} = & \epsilon 
	n^{\lambda}h_{\alpha}{}^{\mu}h_{\beta}{}^{\nu}\nabla_{\nu}n_{\mu}
	\nonumber \\
	= & \epsilon n^{\lambda}K_{\alpha\beta}, \\
\label{eq:coDimOne_A}
	A^{\lambda}{}_{\alpha\beta} = & 
	- a^{\lambda} n_{\alpha}n_{\beta} , \\
\label{eq:coDimOne_Rperp}
	{}^{\perp}R_{\alpha\beta\gamma\delta} = & 0 , 
\end{align} 
	where we have defined $a^{\mu}\equiv n^{\nu}\nabla_{\nu}n^{\mu}$, which
is perpendicular to $n_{\mu}$ so that 
$a^{\mu}h_{\mu}{}^{\lambda}=a^{\lambda}$, and 
$K_{\mu\nu}$ is the standard second fundamental form for codimension
one surfaces. We see that the torsion tensor 
$F^{\lambda}{}_{\alpha\beta}=0$. The projected Ricci tensor components
are
\begin{align} 
	h_{\alpha}{}^{\mu}h_{\beta}{}^{\lambda}R_{\mu\lambda}
	= & 
	-\epsilon h_{\alpha}{}^{\mu}h_{\beta}{}^{\lambda}
	  n^{\nu}\nabla_{\nu}K_{\lambda\mu} 
	+ \epsilon {}^{\parallel}\nabla_{\alpha}a_{\beta}
	- a_{\alpha}a_{\beta} - \epsilon K K_{\alpha\beta} 
	+ {}^{\parallel}R_{\alpha\beta}
	, \\
	n^{\mu}n^{\lambda}R_{\mu\lambda}
	= & 
	  {}^{\parallel}\nabla_{\lambda}a^{\lambda}
	- \epsilon a^{\lambda}a_{\lambda}
	- h^{\mu\rho}n^{\nu}\nabla_{\nu}K_{\rho\mu}
	- K^{\sigma\lambda}K_{\sigma\lambda}
	, \\
	n_{\alpha}{}^{\nu}n^{\rho}R_{\nu\rho}
	= & 
	\epsilon 
	\left(
	{}^{\parallel}\nabla_{\lambda}K^{\lambda}{}_{\alpha}
	- {}^{\parallel}\nabla_{\alpha}K  
	\right)
	, 
\end{align}
from which one can derive the standard $1+(d-1)$ projected Einstein
equations. The projected Ricci scalar is 
\begin{equation}
	R = {}^{\parallel}R 
	+ \epsilon\left(K^2 - K_{\mu\nu}K^{\mu\nu}\right)
	- 2 \epsilon \nabla_{\lambda}
		\left(n^{\lambda}K - a^{\lambda}\right) . 
\end{equation} 
	Here we have defined $K\equiv K_{\mu}{}^{\mu}$, and used the fact that
$n^{\mu}a_{\mu}=0$, so that $h_{\mu}{}^{\alpha}a_{\alpha}=a_{\mu}$. 

\subsection{\label{nullFoliation}Relation between double null 
and codimension two foliations}
	The $m+n$ formalism we have described is capable of 
describing the geometry of double null foliations. 
In a double null foliation, spacetime is foliated by a pair of lightlike 
surfaces, $\Sigma^0$ and $\Sigma^1$, which have the null generators
$l_{\alpha}^{(0)}$ and $l_{\alpha}^{(1)}$, respectively 
\cite{Brady:1995na}. The intersections of the foliations,
$\{\Sigma^0\}\cap\{\Sigma^1\}$ form a spacelike foliation of codimension two,
which we then identify as the foliation $\Sigma$. The transverse space 
$T_p({}^{\perp}\Sigma)$ for each point $p\in\Sigma$ is spanned by the two
null generators $(l^{(0)})^{\alpha}$ and $(l^{(1)})^{\alpha}$. 
We can now define the transverse projection operator as
\begin{equation}
\label{eq:nullTransverseProjection}
	l_{\mu\nu} = l^{(0)}_{\mu}l^{(1)}_{\nu} + l^{(0)}_{\nu}l^{(1)}_{\mu}.
\end{equation}	
	The tangent projector can then be computed from the relation 
$h_{\mu\nu}=g_{\mu\nu}-l_{\mu\nu}$.  

\section{\label{ADMLikeVariables}
	ADM-like variables for $m+n$ spacetime splitting}

	In this section, we set up a coordinate system adapted to the foliation
$(\Sigma_{\bf t})_{{\bf t}\in\mathbb{R}^m}$. We then write down
the tensors 
${}^{\parallel}R_{\alpha\beta\gamma\delta}$,
${}^{\perp}R_{\alpha\beta\gamma\delta}$,
$K^{\lambda}{}_{\alpha\beta}$, and
$A^{\lambda}{}_{\alpha\beta}$ as functions of these coordinates. 
We closely follow the work of \cite{Brady:1995na} in defining
the basis vectors for $T_p(\Sigma)$ and $T_p({}^{\perp}\Sigma)$; see 
also \cite{PoissonBook,2010PhRvD..82l4018H} for similar treatments
of this subject. 

We recall our notation: Greek indices run from $0,...,d-1$, 
lower case Latin indices
run from $0,...,m-1$ and upper case Latin indices from from $m,...,d-1$. 
Einstein summation notation will apply to all different index types. 

\subsection{Coordinate system and metric decomposition}

	We begin by setting up a coordinate system on our manifold $M$ adapted to an $m+n$
spacetime foliation. The coordinates $x^{\alpha}$ of some  
chart of the spacetime manifold $M$
are written as functions of two sets of variables, $\{u^a\}$ and 
$\{\theta^A\}$, $x^{\alpha}\equiv x^{\alpha}(u^a,\theta^A)$.
Derivatives with respect to the variables $u^a$ will be denoted by
$\partial_a\equiv\partial/\partial u^a$, while derivatives with respect
to the variables $\theta^A$ will be denoted by
$\partial_A\equiv\partial/\partial\theta^A$.
	The set $\{\theta^A\}$ are the intrinsic coordinates on the leaf $\Sigma$.
The $\{u^a\}$ are scalar fields, the level sets of which define a congruence
of curves that intersect all the leafs $\Sigma$ of the foliation. In other
words, for the leaf $\Sigma_{\bf t\in\mathbb{R}^m}$, we have
\begin{equation}
	{\bf t} = (u^0,...,u^{m-1})\,.
\end{equation}
We use this
congruence to relate coordinates on each leaf to each other. For example,
in the $1+(d-1)$ formalism ${\bf t} = u^0\equiv t$, the time function. Just as
in the $1+(d-1)$ formalism, we neither assume that the congruence of curves to
be geodesics nor assume that they are orthogonal to the leafs $\Sigma$. 
The tangent vector for
the congruence defined by $u^c$ is denoted by
\begin{equation}
\label{eq:congruenceTangent}
	u_c^{\gamma} \equiv \partial_c x^{\gamma} .
\end{equation} 
	This is to be compared to the $1+(d-1)$ formalism, where the time tangent
vector is often denoted by $t^{\alpha} \equiv \partial_tx^{\alpha}$.

	We now define a coordinate basis on the leaf $\Sigma$ as follows
\begin{equation}
\label{eq:EDef}
	e^{\alpha}_A \equiv \partial_A x^{\alpha}, 
\end{equation}	
	from which we can construct the intrinsic metric on $\Sigma$
\begin{equation}
\label{eq:inducedMetric}
	\gamma_{AB} \equiv g_{\alpha\beta}e^{\alpha}_Ae^{\beta}_B.
\end{equation}
	We will raise/lower capital Latin indices with 
$\gamma_{AB}$ and $\gamma^{AB}$ respectively, where
$\gamma^{AB}$ is the inverse of the induced metric $\gamma_{AB}$. The
metric covariant derivative with respect to $\gamma_{AB}$ will be denoted 
as ${}^{\parallel}{\nabla}_A$. 
At each point $p\in\Sigma$, we can define a basis for
$T_p^*({}^{\perp}\Sigma)$ as follows:
\begin{equation}
\label{eq:NDef}
	n^a_{\alpha} \equiv \partial_{\alpha} u^a\,.
\end{equation}	
	The one-forms $\{n^a_{\alpha}\}$ need not be orthonormal with one another;
we capture this lack of orthonormality with the following symmetric 
inner product matrix
\begin{equation}
	\alpha^{ab} \equiv g^{\alpha\beta}n_{\alpha}^an_{\beta}^b ,
\end{equation} 
	which is symmetric in $a\leftrightarrow b$. As the $\{n^a_{\alpha}\}$ are
form a basis for $T^*_p({}^{\perp}\Sigma)$, $\alpha^{ab}$ is invertible
and we denote its matrix inverse by $\alpha_{ab}$;
$\alpha_{ac}\alpha^{cb}=\delta_a^b$, where $\delta_a^b$ is
the Kronecker delta symbol. 
We emphasize that $\alpha^{ab}$ is \emph{not} an induced metric on the
transverse space ${}^{\perp}\Sigma$, as in general ${}^{\perp}\Sigma$ does 
not integrate to form a submanifold.  
We will formally raise/lower frame indices for the transverse spaces
with the inner product 
matrices $\alpha^{ab}$ and $\alpha_{ab}$, respectively. The spacetime
scalar $\alpha_{ab}$ corresponds to a generalization of the lapse
function $\alpha$ in the $1+(d-1)$ formalism. In particular, in the $1+(d-1)$
formalism we identify 
$\alpha_{00}=-\alpha^2$ and $\alpha^{00}=-\alpha^{-2}$.
The unit normal forms to the leaves $\Sigma$ are computed as follows, 
\begin{equation}
	n_{a\alpha} \equiv \alpha_{ab}\partial_{\alpha}u^b .
\end{equation} 

	We now introduce a generalization of the shift vector.
	With the above definitions in hand, we see that the vectors 
\begin{equation}
\label{eq:orthogonalN}
	\left\{
	u^{\alpha}_a - n^{\alpha}_a
	\right\}_{a=0,...,m-1}
\end{equation}  
	are orthogonal to the one forms $\{n_{\beta}^b\}_{b=0,...,m-1}$. 
From this we 
conclude that we can write the vector $n^{\alpha}_a$ as 
\begin{equation}
\label{eq:NDef2}
	n^{\alpha}_a \equiv u^{\alpha}_a - \beta^{\alpha}_a ,
\end{equation}
	where we have defined the shift vectors $\{\beta^{\alpha}_a\}$, which are
orthogonal to the one forms $n_{a\alpha}$; 
i.e. $n_{a\alpha}\beta^{\alpha}_b=0$. The shift vectors 
$\{\beta^{\alpha}_a\}_{a=0,...,m-1}$ are
a direct generalization of the shift vector $\beta^{\alpha}$ in the $1+(d-1)$
formalism. 
. 

We next derive some useful relations for $e^{\alpha}_A$ and $n^{\alpha}_a$.
	The relations Eq.~\eqref{eq:EDef} and \eqref{eq:NDef2} imply that in the
coordinates $(u^a,\theta^A)$ we have 
\begin{align}
\label{eq:EFrameVal}
	e^{\alpha}_A & \overset{*}{=} \delta^{\alpha}_A , \\
\label{eq:NFrameVal}
	n^{\alpha}_a & \overset{*}{=} \delta^{\alpha}_a 
		- \beta^A_a\delta^{\alpha}_A,
\end{align}
	where the $\delta$ is the Kronecker delta symbol and 
$\overset{*}{=}$ means that this only holds in the specific coordinate
choice $\{(u^a,\theta^A)\}$. We only use the symbol 
$\overset{*}{=}$ in this section; in the
Sections I-V we work with the coordinate choices defined by 
Eqs.~\eqref{eq:EFrameVal} and~\eqref{eq:NFrameVal}.
In the $1+(d-1)$ formalism  the equivalent
coordinate choice would be $\{t,x^i\}$, where the $\{x^i\}_{i=1,2,3}$ 
are the three
spatial directions. We see that in this basis the shift vectors have nonzero
components only on their last $n$ indices: 
$\beta^{\alpha}_a\overset{*}{=}(0,...,0,\beta^A_a)$. 
From Eqs.~\eqref{eq:EFrameVal} and~\eqref{eq:NFrameVal}
we conclude that the frame vectors $e^{\alpha}_A$ are Lie transported
along each of the congruences defined by the level sets of 
the functions $u^c$ 
\begin{equation}
\label{eq:UELieDerivative}
	\pounds_{u^{\gamma}_c}e^{\alpha}_A = 0 .
\end{equation} 
	Since this expression is tensorial, it holds in any coordinate system.
Other useful tensorial relations we can derive from the above expressions are 
\begin{align}
\label{eq:ENLieDerivative}
	\pounds_{e^{\gamma}_C}n^a_{\alpha} 
	= e^{\gamma}_C\nabla_{\gamma}n^a_{\alpha} 
	+ n^a_{\gamma}\nabla_{\alpha}e^{\gamma}_C= & 0 , \\
\label{eq:EELieDerivative}
	\pounds_{e^{\gamma}_C}e^{\alpha}_A 
	= e^{\gamma}_C\nabla_{\gamma}e^{\alpha}_A
	- e^{\gamma}_A\nabla_{\gamma}e^{\alpha}_C = & 0 , \\
\label{eq:NNLieDerivative}
	\pounds_{n^{\alpha}_a}n^{\gamma}_b 
	= n^{\alpha}_a\nabla_{\alpha}n^{\gamma}_b
	- n^{\beta}_b\nabla_{\beta}n^{\gamma}_a = & -\mathcal{F}^{\gamma}_{ab} , 
\end{align}
	where we have defined the transverse torsion spacetime \emph{vector} 
$\mathcal{F}^{\gamma}_{ab}$ to be
\begin{align}
\label{Fabvector}
	\mathcal{F}^{\gamma}_{ab} \equiv & 
	  \partial_a\beta^{\gamma}_b 
	- \partial_b\beta_a^{\gamma} 
	+ \beta^D_b\partial_D\beta^{\gamma}_a 
	- \beta^D_a\partial_D\beta^{\gamma}_b . 
\end{align} 
	The vector $\mathcal{F}^{\gamma}_{ab}$ is orthogonal to the forms 
$n_{\gamma}^c$ 
\begin{equation}
\label{eq:FOrthogonalN}
	n_{\gamma}^c\mathcal{F}^{\gamma}_{ab} = 0 .
\end{equation}
	As this expression is tensorial it holds in general coordinate system. 
In the adapted basis
$\{(u^a,\theta^A)\}$ we may write $\mathcal{F}^{\gamma}_{ab}
=e^{\gamma}_C\mathcal{F}^C_{ab} \overset{*}{=} 
\mathcal{F}^C_{ab}$ to reflect this fact.	

We now see how the metric is $m+n$ decomposed. We begin by decomposing
	the differential $dx^{\alpha}$ into terms tangent
and transverse to the leaf $\Sigma$ \cite{Brady:1995na}
\begin{equation}
\label{eq:dxDecomp}
	dx^{\alpha} = n_a^{\alpha}du^a 
	+ e^{\alpha}_A\left(d\theta^A + \beta^A_adu^a\right) .
\end{equation} 
	From which the spacetime line element can be written as	
\begin{equation}
\label{coordinateBasisLine}
	ds^2 = \alpha_{ab}du^adu^b + \gamma_{AB}
	\left(d\theta^A + \beta^A_adu^a\right)
	\left(d\theta^B + \beta^B_bdu^b\right) .	
\end{equation}
	We note that with the spacetime line element
Eq.~\eqref{coordinateBasisLine} the 
 metric determinant factorizes as follows
\begin{equation}
\label{eq:spacetimeDet}
	\mathrm{det}\left(g_{\mu\nu}\right) = 
	\mathrm{det}\left(\alpha_{ab}\right)
	\mathrm{det}\left(\gamma_{AB}\right) .
\end{equation}
	We compare Eq.~\eqref{eq:spacetimeDet} to the case in the $1+(d-1)$
formalism, where $\mathrm{det}(g)=\alpha^2\mathrm{det}(\gamma_{ij})$. 
	Furthermore, we have the following relations
\begin{align}
	h_{\alpha\beta} = & \gamma_{AB}e^A_{\alpha}e^B_{\beta} , \\
	l_{\alpha\beta} = & \alpha_{ab}n^a_{\alpha}n^b_{\beta} ,
\end{align}
	so that the metric can be written as follows 
(see, for example \cite{Brady:1995na,PhysRevD.56.6284} for similar
presentations of the metric tensor)
\begin{equation}
	g_{\alpha\beta} = 
		\alpha_{ab}n^a_{\alpha}n^b_{\beta} 
		+ \gamma_{AB}e^A_{\alpha}e^B_{\beta} 
	.
\end{equation}
		
\subsection{\label{RewritingCurvatureTerms}
	Rewriting curvature terms in ADM-like variables} 	
	In this section, we compute the components of
$K_{\gamma\alpha\beta}$, 
${}^{\parallel}R_{\alpha\beta\gamma\delta}$,
$A_{\gamma\alpha\beta}$, and
${}^{\perp}R_{\alpha\beta\gamma\delta}$
in the adapted basis $\{(u^a,\theta^A)\}$, i.e. when 
the relations Eqs.~\eqref{eq:EFrameVal} 
and \eqref{eq:NFrameVal} hold. 
The curvature terms $K_{\gamma\alpha\beta}$
and ${}^{\parallel}R_{\alpha\beta\gamma\delta}$
have direct analogues in the $1+(d-1)$ formalism, and can be computed
as functions of the metric Eq.~\eqref{coordinateBasisLine} in a 
way analogous to what is done in the $1+(d-1)$ formalism. We have 
found a greater variety of functional forms for the curvature terms  
$A_{\gamma\alpha\beta}$ and 
${}^{\perp}R_{\alpha\beta\gamma\delta}$ that have been presented
in the literature. We recall that $\alpha_{ab}$ is generally not the induced
metric for any submanifold, as the transverse space ${}^{\perp}\Sigma$
can only integrate to 
form a manifold in factorizable spacetimes
(see Appendix \ref{factorizableSpacetime}). 

\subsubsection{Computing $K_{\gamma\alpha\beta}$}
	We first compute $K_{\gamma\alpha\beta}$. We have
\begin{align}
	K_{\gamma\alpha\beta} 
	= & h_{\alpha}{}^{\mu}h_{\beta}{}^{\nu}\nabla_{\nu}l_{\mu\gamma} 
	 \nonumber \\
	= & e^A_{\alpha}e^B_{\beta}n^c_{\gamma}
		\left(e^{\mu}_Ae^{\nu}_B\nabla_{\nu}n_{c\mu}\right)
	 \nonumber \\
	= & e^A_{\alpha}e^B_{\beta}n^c_{\gamma}
		\frac{1}{2}\left(
		e^{\mu}_Ae^{\nu}_B\pounds_{n_c^{\gamma}}g_{\mu\nu}
		\right)
	 \nonumber \\
	\equiv & e^A_{\alpha}e^B_{\beta}n^c_{\gamma}
		\mathcal{K}_{cAB} .
\end{align}
	We now rewrite the Lie derivative of $n^{\gamma}_c$ in terms of covariant
derivatives acting on the shift vectors $\beta^{\gamma}_c$ 
and the Lie derivative of $u^{\gamma}_c$. We compute	
\begin{align}
	\pounds_{u^{\gamma}_c}\gamma_{AB} 
	= & \pounds_{u^{\gamma}_c}
		\left(e^{\alpha}_Ae^{\beta}_Bg_{\alpha\beta}\right) 
	 \nonumber \\
	= & e^{\alpha}_Ae^{\beta}_B\pounds_{u^{\gamma}_c}g_{\alpha\beta} 
	 \nonumber \\
	= & e^{\alpha}_Ae^{\beta}_B
		\left(\nabla_{\alpha}u_{c\beta} + \nabla_{\beta}u_{c\alpha}\right) 
	 \nonumber \\
	= & e^{\alpha}_Ae^{\beta}_B\left(
		\nabla_{\alpha}n_{c\beta}+\nabla_{\beta}n_{c\alpha}
		+ \nabla_{\alpha}\beta_{c\beta}+\nabla_{\beta}\beta_{c\alpha}
		\right) 
	 \nonumber \\
	= & e^{\alpha}_Ae^{\beta}_B\pounds_{n^{\gamma}_c}g_{\alpha\beta}
		+ {}^{\parallel}\nabla_A\beta_{cB} + {}^{\parallel}\nabla_B\beta_{cA}
		 .
\end{align} 
	From this we conclude that 
\begin{align}
\label{eq:extrinsicKBasis}
	K_{\gamma\alpha\beta} 
	= & e^A_{\alpha}e^B_{\beta}n^c_{\gamma}\mathcal{K}_{cAB} 
	, \nonumber \\ 
	\mathcal{K}_{cAB} \equiv & \frac{1}{2}
		\left(\pounds_{u^{\gamma}_c}\gamma_{AB}
		-{}^{\parallel}\nabla_A\beta_{cB}
		-{}^{\parallel}\nabla_B\beta_{cA}\right) .
\end{align}
	Recall that lower case Latin letters act as labels, so that
${}^{\parallel}\nabla_A\beta_{cB} 
	= \partial_A\beta_{cB} - \Gamma^C{}_{AB}\beta_{cC}$,
where $\Gamma^C{}_{AB}$ is defined by Eq.~\eqref{eq:DefGammaCAB}. Also
note that as $\gamma_{AB}$ is a spacetime scalar, in the coordinate adapted
basis we have $\pounds_{u^{\gamma}_c}\gamma_{AB}=\partial_c\gamma_{AB}$. This
is to be compared to the $1+(d-1)$ spacetime splitting formalism, where
instead one has $\pounds_{t^{\alpha}}\gamma_{ij}\equiv\partial_t\gamma_{ij}$,
and there is only one shift vector $\beta^i$.

\subsubsection{Computing 
	${}^{\parallel}\mathcal{R}_{\alpha\beta\gamma\delta}$}

Next, we compute ${}^{\parallel}\mathcal{R}_{\alpha\beta\gamma\delta}$.
	The connection coefficients for the induced
covariant derivative on $\Sigma$ is computed as follows:
\begin{align}
\label{eq:DefGammaCAB}
	\Gamma_{CAB} 
	\equiv & e^{\beta}_Be_{C\alpha}\nabla_{\beta}e^{\alpha}_A
	 \nonumber \\
	= & \frac{1}{2}\left(\partial_A\gamma_{BC} + \partial_B\gamma_{AC}
			- \partial_C\gamma_{AB}\right) . 
\end{align} 
	Note that $\Gamma^C{}_{AB}=\gamma^{CD}\Gamma_{DAB}$. We can now compute 
${}^{\parallel}R_{\alpha\beta\gamma\delta}$ in
terms of contractions and derivatives of the connection $\Gamma_{CAB}$. 
\begin{align}
	{}^{\parallel}R_{\alpha\beta\gamma}{}^{\delta}e^D_{\delta} = & 
2 h_{[\alpha}{}^{\mu}h_{\beta]}{}^{\nu}h_{\gamma}{}^{\lambda}
\nabla_{\mu}\left(h_{\nu}{}^{\eta}h_{\lambda}{}^{\rho}
\nabla_{\eta}e^D_{\rho}\right) 
	 \nonumber \\
	= & 2 e^A_{[\alpha}e^B_{\beta]}e^C_{\gamma}\left(
e^{\mu}_A\nabla_{\mu}\Gamma_C{}^D{}_B + \Gamma_I{}^D{}_B\Gamma_C{}^I{}_A 
+ \Gamma_C{}^D{}_I\Gamma_B{}^I{}_A\right) 
	 \nonumber \\
	\overset{*}{=} & 2 e^A_{[\alpha}e^B_{\beta]}e^C_{\gamma}
	\left(
	- \partial_A\Gamma^D{}_{CB}
	+ \Gamma^D{}_{IB}\Gamma^I{}_{CA}
	\right) . 
\end{align}	
	To obtain the third line we used the property
$\Gamma_B{}^I{}_A=-\Gamma^I{}_{BA}$. We also used the fact
that $\gamma_{CAB}$ is a spacetime scalar in $M$, and in our
coordinate basis $e^{\mu}_A\overset{*}{=}\delta^{\mu}_A$ so
that $e^{\mu}_A\nabla_{\mu}\Gamma_{CAB}\overset{*}{=}\partial_A\Gamma_{CAB}$.
We conclude that
\begin{align}
\label{eq:ParallelRiemannDecompBasis}
	{}^{\parallel}R_{\alpha\beta\gamma\delta} 
	\overset{*}{=} & e^A_{\alpha}e^B_{\beta}e^C_{\gamma}e^D_{\delta}
	{}^{\parallel}\mathcal{R}_{ABCD} , 
\end{align}
	where
\begin{align}
\label{eq:ParallelRiemannBasis}	
{}^{\parallel}\mathcal{R}^D{}_{CAB} 
	\equiv & 
	\partial_A\Gamma^D{}_{CB} - \partial_B\Gamma^D{}_{CA} 
	+ \Gamma^D{}_{IA}\Gamma^I{}_{CB} - \Gamma^D{}_{IB}\Gamma^I{}_{CA} .	
\end{align}

\subsubsection{Computing $A_{\gamma\alpha\beta}$}
	Let us next compute $A_{\gamma\alpha\beta}$.
	We define the quantity 
$\mathcal{A}^C_{ab}\equiv n^{\alpha}_an^{\beta}_b\nabla_{\beta}e_{\alpha}^C$,
so that
$A_{\gamma\alpha\beta}
= e_{C\gamma}n^a_{\alpha}n^b_{\beta} \mathcal{A}^C_{ab}$. The antisymmetric
part of $\mathcal{A}^C_{ab}$ (i.e. the transverse torsion) is 
\begin{align}
	\mathcal{A}^C_{[ab]} = & - 
		\frac{1}{2}e^C_{\gamma}\left[n_b, n_a\right]^{\gamma}
	 \nonumber \\
	= & - \frac{1}{2}e^C_{\gamma}\mathcal{F}^{\gamma}_{ab}	,
\end{align}	
	where we obtained the second line using
Eq.~\eqref{eq:NNLieDerivative}. The symmetric part of $\mathcal{A}_C^{ab}$ is
\begin{align}
	\mathcal{A}_C^{(ab)} = &
	\frac{1}{2}n_{\alpha}^an_{\beta}^b\pounds_{e^{\gamma}_C}g^{\alpha\beta} 
	 \nonumber \\
	= & \frac{1}{2}\pounds_{e^{\gamma}_C}\alpha^{ab} .
\end{align}
	The second line holds as a result of Eq.~\eqref{eq:ENLieDerivative} and
the definition of $\alpha^{ab}$. Using 
Eqs.~\eqref{eq:EFrameVal} and~\eqref{eq:NFrameVal}, we conclude that
\begin{equation}
\label{eq:extrinsicABasis}
	\mathcal{A}_C^{ab}
	\overset{*}{=} \frac{1}{2}
	\left(
	\partial_C\alpha^{ab} 
	- \alpha^{ac}\alpha^{bd}\gamma_{CD}\mathcal{F}^D_{cd}	
	\right) .
\end{equation} 

\subsubsection{Computing
${}^{\perp}\mathcal{R}_{\alpha\beta\gamma\delta}$}
We now compute ${}^{\perp}\mathcal{R}_{\alpha\beta\gamma\delta}$.
	We define the quantity
\begin{align}
\label{eq:DefOmegacab}
	\Omega_{cab} \equiv & n^{\beta}_bn_{c\alpha}\nabla_{\beta}n^{\alpha}_a 
	 \nonumber \\
	= & \frac{1}{2}\left(
	  n_a^{\mu}\partial_{\mu}\alpha_{bc} 
	+ n_b^{\mu}\partial_{\mu}\alpha_{ac} 
	- n_c^{\mu}\partial_{\mu}\alpha_{ab} 
	\right) . 
\end{align}
	To derive the second line of the above we used Eq.~\eqref{eq:FOrthogonalN}.
Note that in the coordinate adapted basis, Eq.~\eqref{eq:NFrameVal} we have
$n^{\mu}_a\partial_{\mu}\Omega_{kij} \overset{*}{=} 
	\partial_a\Omega_{kij}-\beta^A_a\partial_A\Omega_{kij}$. Similarly to
$\Gamma_{CAB}$, whose first index can be raised with $\gamma^{CD}$, we can
raise the first index of $\Omega_{cab}$ with $\alpha^{cd}$,
$\Omega^c{}_{ab}=\alpha^{cd}\Omega_{dab}$.
We now look at 
\begin{align}
\label{eq:interPRiemann01}
	{}^{\perp}R_{\alpha\beta\gamma}{}^{\delta}n_{\delta}^d = & 
2 l_{[\alpha}{}^{\mu}l_{\beta]}{}^{\nu}l_{\gamma}{}^{\lambda}
\nabla_{\mu}\left(l_{\nu}{}^{\eta}l_{\lambda}{}^{\rho}
\nabla_{\eta}n^d_{\rho}\right) 
	- 2A^{\lambda}{}_{[\alpha\beta]}l_{\gamma}{}^{\delta}
		\nabla_{\lambda}n^d_{\delta} . 
\end{align} 
	We first focus on the last term of this expression. Using 
Eq.~\eqref{eq:ENLieDerivative}, we see that 
\begin{align}
	- 2A^{\lambda}{}_{[\alpha\beta]}l_{\gamma}{}^{\delta}
		\nabla_{\lambda}n^d_{\delta} 
	= 2 n_{a\alpha}n_{b\beta}n_{c\gamma}\gamma^{CD}
		\mathcal{A}_C^{[ab]}\mathcal{A}_D^{dc}  .
\end{align}
	The first term of Eq.~\eqref{eq:interPRiemann01} is	
\begin{align}
	2 l_{[\alpha}{}^{\mu}l_{\beta]}{}^{\nu}l_{\gamma}{}^{\lambda}
\nabla_{\mu}\left(l_{\nu}{}^{\eta}l_{\lambda}{}^{\rho}
\nabla_{\eta}n^d_{\rho}\right)
	= & 2n_{[\alpha}^an_{\beta]}^bn_{\gamma}^c
	\left(
	n^{\mu}_a\nabla_{\mu}\Omega_c{}^d{}_b + \Omega_b{}^i{}_a\Omega_c{}^d{}_i
	+ \Omega_c{}^j{}_a\Omega_j{}^d{}_b
	\right)
	 \nonumber \\
	\overset{*}{=} & 2n_{[\alpha}^an_{\beta]}^bn_{\gamma}^c
	\left(
	n^{\mu}_b\partial_{\mu}\Omega^d{}_{ac} + \Omega^d{}_{ib}\Omega^i{}_{ac}
	\right) . 
\end{align}
	To calculate the second line we have made use of the identities
$\Omega_a{}^c{}_b=-\Omega^c{}_{ab}$ and $\Omega_{c[ab]}=0$, which follow
from Eq.~\eqref{eq:DefOmegacab}. We conclude that
\begin{align}
\label{eq:PerpRiemannDecompBasis}
	{}^{\perp}R_{\alpha\beta\gamma\delta} 
	\overset{*}{=} n^a_{\alpha}n^b_{\beta}n^c_{\gamma}n^d_{\delta}
	\left(
	{}^{\perp}\mathcal{R}_{abcd} 
	+ 2 \alpha_{ai}\alpha_{bj}\alpha_{ck}\alpha_{dl}\gamma^{CD}
	\mathcal{A}^{[ij]}_C\mathcal{A}^{lk}_D
	\right) ,
\end{align}
	where
\begin{equation}
\label{eq:PerpRiemannBasis}	
	{}^{\perp}\mathcal{R}^d{}_{cab}
	\equiv n^{\mu}_a\partial_{\mu}\Omega^d{}_{cb} 
	- n^{\mu}_b\partial_{\mu}\Omega^d{}_{ca}
	+ \Omega^d{}_{ia}\Omega^i{}_{cb}
	- \Omega^d{}_{ib}\Omega^i{}_{ca} 
	.
\end{equation}

\subsubsection{\label{ProjEH}Projected Einstein-Hilbert action}
	Having the above results at hand, we now rewrite the Einstein-Hilbert 
action in $d$ dimensional spacetime 
\begin{equation}
	S = \frac{1}{2}\int d^{d}x\sqrt{-g}R ,
\end{equation}
	in an $m+n$ decomposition. Using Eqs.
\eqref{eq:spacetimeDet}, \eqref{eq:extrinsicKBasis}, 
\eqref{eq:extrinsicABasis}, 
\eqref{eq:ParallelRiemannDecompBasis}, and 
\eqref{eq:PerpRiemannDecompBasis},
we have
\begin{align}
\label{eq:EHActionADM}
	S \overset{*}{=} \int d^mud^n\theta 
	\sqrt{\alpha}\sqrt{\gamma}
	\Big( &  
	{}^{\parallel}\mathcal{R}
	+ \alpha^{cd}\gamma^{AB}\gamma^{CD}
		\left(
	  \mathcal{K}_{cAB}\mathcal{K}_{dCD} 
	- \mathcal{K}_{cAC}\mathcal{K}_{dBD} 
		\right)
	\nonumber \\ &  
	+ {}^{\perp}\mathcal{R}
	+ \gamma^{CD}\alpha_{ab}\alpha_{cd}
		\left(
	  \mathcal{A}_C^{ab}\mathcal{A}_D^{cd} 
	- \mathcal{A}_C^{ac}\mathcal{A}_D^{bd} 
		\right)
	- 2 \nabla_{\lambda}\left(K^{\lambda}+A^{\lambda}\right)
	\Big) .
\end{align} 
	We can recover the complete Einstein equations
by varying the Einstein-Hilbert action, Eq.~\eqref{eq:EHActionADM} with
respect to 
$\{\alpha_{ab}\}_{a,b=0,...,m-1}$, 
$\{\beta^{\alpha}_a\}_{a=0,..,m-1}$, and
$\gamma_{AB}$. This is to be compared to the $1+(d-1)$ formalism, where one
varies the Einstein-Hilbert action with respect to $\alpha$, $\beta^i$, 
and $\gamma_{ij}$, with $\alpha$ and $\beta^i$ acting as constraint
variables. Care must be taken when varying Eq.~\eqref{eq:EHActionADM}
as in general  
$\alpha_{ab}$ cannot be treated as a metric so there
is in general no well defined notion of a metric compatible connection
for $\alpha_{ab}$, and we 
have relations such as ${}^{\parallel}\nabla_A\alpha_{ab}\neq0$. For
a general spacetime with no symmetries, a potentially more
straightforward approach to finding the Einstein equations in 
the $m+n$ formalism is to contract the projected Riemann 
tensor relations, 
Eqs.~\eqref{eq:hhhhR}, \eqref{eq:llllR}, 
\eqref{eq:hhllR}, \eqref{eq:hhhlR}, and
\eqref{eq:lllhR} to obtain the projected Ricci tensor relations. 

\subsection{\label{factorizableSpacetime}
	$m+n$ splitting in a factorizable spacetime}
	In a factorizable spacetime the spacetime manifold can be written
globally as $M=\Sigma^{(1)}\times\Sigma^{(2)}$, where both
$\Sigma^{(i)}$ are submanifolds of $M$. In a factorizable spacetime,
we see that we can think of either a family of submanifolds
$\{\Sigma^{(2)}_{\bf t}\}$ foliating $M$, indexed by coordinates
on $\Sigma^{(1)}$, or vice-versa. 	
In the context of general relativity in four dimensions, an 
important class of a factorizable spacetimes are spherically
symmetric spacetimes, which take the form $M=M^2 \times S^2$, where
$M^2$ is a two dimensional Lorentzian manifold and $S^2$ is the two
sphere. 
	In factorizable spacetimes, we can choose an adapted basis
to this foliation structure so that the shift vectors 
$\{\beta^{\alpha}_a\}_{a=0,...,m-1}$ all vanish, so that the metric
can be written as
\begin{equation}
\label{eq:FactorizableMetric}
	ds^2 = \alpha_{ab}du^adu^b + \gamma_{AB}d\theta^Ad\theta^B .
\end{equation}
	Unlike in the general $m+n$ decomposition, 
We can introduce a two metric compatible derivative for
the submanifolds $\Sigma^{(1)}$ and $\Sigma^{(2)}$, which we denote
by ${}^{\perp}\nabla_a$ and ${}^{\parallel}\nabla_A$, 
respectively. We see that $\Omega_{cab}$ takes on the role of the connection 
of the
submanifold $(\Sigma^{(1)},{}^{\perp}\nabla_a,\alpha_{ab})$. Writing down 
formulas for ${}^{\perp}\mathcal{R}^a{}_{bcd}$ and $\mathcal{A}_{Cab}$ become
much simpler than in the general $m+n$ case as the shift vectors all
vanish; in particular the directional 
derivatives along $n^{\alpha}_a$ become derivatives in the coordinate
$u^a$; $n^{\alpha}_a\partial_{\alpha}\to\partial_a$.

\section{\label{SVTSpherical}Scalar, vector, and tensor spherical harmonics}
	In this section, we review the properties of the scalar, vector, and 
tensor spherical harmonics. We work on the two sphere $S^2$, with the round 
metric $\Omega_{AB}$ and metric compatible covariant derivative 
$D_A$: $(S^2,\Omega_{AB},D_A)$.  

We begin with the scalar spherical harmonics. Such harmonics satisfy the 
following eigenvalue equation:
\begin{equation}
	\left\{\Omega^{AB}D_AD_B + l(l+1)\right\}Y^{lm} = 0 . 
\end{equation} 
	The scalar spherical harmonics form an orthogonal basis for functions in 
$S_2$. We choose the following normalization for $Y^{lm}$ 
\begin{equation}
	\int d^2\Omega Y^{lm}Y^{l^{\prime}m^{\prime}} 
	= \delta_{ll^{\prime}}\delta_{mm^{\prime}} .
\end{equation}

	Next, we discuss vector spherical harmonics. The axial and polar 
spherical harmonics respectively are 
\begin{align}
	E^{lm}_A = & D_AY^{lm} , \quad 
	B^{lm}_A =  \epsilon_A{}^BD_BY^{lm} . 
\end{align}
	Note that divergence of $B^{lm}_A$ is zero, $D_AB^A_{lm} = 0$. 
The vector spherical harmonics satisfy the following eigenvalue equation: 
\begin{equation}
	\left\{\Omega^{AB}D_AD_B + \left[-1 + l(l+1)\right]\right\}V^{lm}_C = 0 ,
\end{equation} 
	where $V^{lm}_C$ is either $E^{lm}_C$ or $B^{lm}_C$. The vector spherical 
harmonics form an orthonormal basis for functions in $S_2$. The vector 
spherical harmonics are orthogonal to one another, and are normalized to obey 
\begin{equation}
	\int d^2\Omega \; \Omega^{AB}V^{lm}_{A}V_B^{l^{\prime}m^{\prime}} 
	= l(l+1) \delta_{ll^{\prime}}\delta_{mm^{\prime}} .
\end{equation}

	Finally, we introduce tensor spherical harmonics. We define such 
harmonics to be traceless; this choice follows, for example Poisson and 
Martel~\cite{Martel:2005ir}, but not Regge and Wheeler
~\cite{PhysRev.108.1063}. The traceless axial and polar tensor spherical 
harmonics respectively are
\begin{align}
	E^{lm}_{AB} = & D_{(A}E^{lm}_{B)} + \frac{l(l+1)}{2}\Omega_{AB}Y^{lm} ,\\
	B^{lm}_{AB} = & D_{(A}B_{B)}^{lm} .
\end{align}
	The trace can be captured with $Y^{lm}\Omega_{AB}$,
which behaves as a scalar under rotations. The tensor spherical 
harmonics satisfy the following eigenvalue equation:
\begin{equation}
	\left\{\Omega^{AB}D_AD_B 
	+ \left[ -2 + l(l+1) \right]\right\}T_{CD}^{lm} = 0 , 
\end{equation}
	where $T_{CD}^{lm}$ is either $E^{lm}_{CD}$ or $B^{lm}_{CD}$. The trace 
term $Y^{lm}\Omega_{AB}$ has the scalar spherical harmonic eigenvalue 
$l(l+1)$. Finally, the tensor spherical harmonics satisfy the following 
orthogonality relation 
\begin{equation} 
	\int d^2\Omega \; \Omega^{AB}\Omega^{CD}T^{lm}_{AC}
T^{l^{\prime}m^{\prime}}_{BD} = \frac{1}{2}l(l+1)
\left[l(l+1)-2\right]\delta_{ll^{\prime}}\delta_{mm^{\prime}} . 
\end{equation}

\bibliography{/home/jripley/Documents/Research/globalbib}

\begin{thebibliography}{43}%
\makeatletter
\providecommand \@ifxundefined [1]{%
 \@ifx{#1\undefined}
}%
\providecommand \@ifnum [1]{%
 \ifnum #1\expandafter \@firstoftwo
 \else \expandafter \@secondoftwo
 \fi
}%
\providecommand \@ifx [1]{%
 \ifx #1\expandafter \@firstoftwo
 \else \expandafter \@secondoftwo
 \fi
}%
\providecommand \natexlab [1]{#1}%
\providecommand \enquote  [1]{``#1''}%
\providecommand \bibnamefont  [1]{#1}%
\providecommand \bibfnamefont [1]{#1}%
\providecommand \citenamefont [1]{#1}%
\providecommand \href@noop [0]{\@secondoftwo}%
\providecommand \href [0]{\begingroup \@sanitize@url \@href}%
\providecommand \@href[1]{\@@startlink{#1}\@@href}%
\providecommand \@@href[1]{\endgroup#1\@@endlink}%
\providecommand \@sanitize@url [0]{\catcode `\\12\catcode `\$12\catcode
  `\&12\catcode `\#12\catcode `\^12\catcode `\_12\catcode `\%12\relax}%
\providecommand \@@startlink[1]{}%
\providecommand \@@endlink[0]{}%
\providecommand \url  [0]{\begingroup\@sanitize@url \@url }%
\providecommand \@url [1]{\endgroup\@href {#1}{\urlprefix }}%
\providecommand \urlprefix  [0]{URL }%
\providecommand \Eprint [0]{\href }%
\providecommand \doibase [0]{http://dx.doi.org/}%
\providecommand \selectlanguage [0]{\@gobble}%
\providecommand \bibinfo  [0]{\@secondoftwo}%
\providecommand \bibfield  [0]{\@secondoftwo}%
\providecommand \translation [1]{[#1]}%
\providecommand \BibitemOpen [0]{}%
\providecommand \bibitemStop [0]{}%
\providecommand \bibitemNoStop [0]{.\EOS\space}%
\providecommand \EOS [0]{\spacefactor3000\relax}%
\providecommand \BibitemShut  [1]{\csname bibitem#1\endcsname}%
\let\auto@bib@innerbib\@empty
\bibitem [{\citenamefont {Regge}\ and\ \citenamefont
  {Wheeler}(1957)}]{PhysRev.108.1063}%
  \BibitemOpen
  \bibfield  {author} {\bibinfo {author} {\bibfnamefont {T.}~\bibnamefont
  {Regge}}\ and\ \bibinfo {author} {\bibfnamefont {J.~A.}\ \bibnamefont
  {Wheeler}},\ }\href {\doibase 10.1103/PhysRev.108.1063} {\bibfield  {journal}
  {\bibinfo  {journal} {Phys. Rev.}\ }\textbf {\bibinfo {volume} {108}},\
  \bibinfo {pages} {1063} (\bibinfo {year} {1957})}\BibitemShut {NoStop}%
\bibitem [{\citenamefont {Vishveshwara}(1970)}]{PhysRevD.1.2870}%
  \BibitemOpen
  \bibfield  {author} {\bibinfo {author} {\bibfnamefont {C.~V.}\ \bibnamefont
  {Vishveshwara}},\ }\href {\doibase 10.1103/PhysRevD.1.2870} {\bibfield
  {journal} {\bibinfo  {journal} {Phys. Rev. D}\ }\textbf {\bibinfo {volume}
  {1}},\ \bibinfo {pages} {2870} (\bibinfo {year} {1970})}\BibitemShut
  {NoStop}%
\bibitem [{\citenamefont {Zerilli}(1970)}]{PhysRevLett.24.737}%
  \BibitemOpen
  \bibfield  {author} {\bibinfo {author} {\bibfnamefont {F.~J.}\ \bibnamefont
  {Zerilli}},\ }\href {\doibase 10.1103/PhysRevLett.24.737} {\bibfield
  {journal} {\bibinfo  {journal} {Phys. Rev. Lett.}\ }\textbf {\bibinfo
  {volume} {24}},\ \bibinfo {pages} {737} (\bibinfo {year} {1970})}\BibitemShut
  {NoStop}%
\bibitem [{\citenamefont {Moncrief}(1974)}]{MONCRIEF1974323}%
  \BibitemOpen
  \bibfield  {author} {\bibinfo {author} {\bibfnamefont {V.}~\bibnamefont
  {Moncrief}},\ }\href {\doibase
  http://dx.doi.org/10.1016/0003-4916(74)90173-0} {\bibfield  {journal}
  {\bibinfo  {journal} {Annals of Physics}\ }\textbf {\bibinfo {volume} {88}},\
  \bibinfo {pages} {323 } (\bibinfo {year} {1974})}\BibitemShut {NoStop}%
\bibitem [{\citenamefont {Gerlach}\ and\ \citenamefont
  {Sengupta}(1979)}]{PhysRevD.19.2268}%
  \BibitemOpen
  \bibfield  {author} {\bibinfo {author} {\bibfnamefont {U.~H.}\ \bibnamefont
  {Gerlach}}\ and\ \bibinfo {author} {\bibfnamefont {U.~K.}\ \bibnamefont
  {Sengupta}},\ }\href {\doibase 10.1103/PhysRevD.19.2268} {\bibfield
  {journal} {\bibinfo  {journal} {Phys. Rev. D}\ }\textbf {\bibinfo {volume}
  {19}},\ \bibinfo {pages} {2268} (\bibinfo {year} {1979})}\BibitemShut
  {NoStop}%
\bibitem [{\citenamefont {Gerlach}\ and\ \citenamefont
  {Sengupta}(1980)}]{PhysRevD.22.1300}%
  \BibitemOpen
  \bibfield  {author} {\bibinfo {author} {\bibfnamefont {U.~H.}\ \bibnamefont
  {Gerlach}}\ and\ \bibinfo {author} {\bibfnamefont {U.~K.}\ \bibnamefont
  {Sengupta}},\ }\href {\doibase 10.1103/PhysRevD.22.1300} {\bibfield
  {journal} {\bibinfo  {journal} {Phys. Rev. D}\ }\textbf {\bibinfo {volume}
  {22}},\ \bibinfo {pages} {1300} (\bibinfo {year} {1980})}\BibitemShut
  {NoStop}%
\bibitem [{\citenamefont {Chandrasekhar}(2002)}]{Chandrasekhar:579245}%
  \BibitemOpen
  \bibfield  {author} {\bibinfo {author} {\bibfnamefont {S.}~\bibnamefont
  {Chandrasekhar}},\ }\href {https://cds.cern.ch/record/579245} {\emph
  {\bibinfo {title} {{The mathematical theory of black holes}}}},\ Oxford
  classic texts in the physical sciences\ (\bibinfo  {publisher} {Oxford Univ.
  Press},\ \bibinfo {address} {Oxford},\ \bibinfo {year} {2002})\BibitemShut
  {NoStop}%
\bibitem [{\citenamefont {Martel}\ and\ \citenamefont
  {Poisson}(2005)}]{Martel:2005ir}%
  \BibitemOpen
  \bibfield  {author} {\bibinfo {author} {\bibfnamefont {K.}~\bibnamefont
  {Martel}}\ and\ \bibinfo {author} {\bibfnamefont {E.}~\bibnamefont
  {Poisson}},\ }\href {\doibase 10.1103/PhysRevD.71.104003} {\bibfield
  {journal} {\bibinfo  {journal} {Phys. Rev.}\ }\textbf {\bibinfo {volume}
  {D71}},\ \bibinfo {pages} {104003} (\bibinfo {year} {2005})},\ \Eprint
  {http://arxiv.org/abs/gr-qc/0502028} {arXiv:gr-qc/0502028 [gr-qc]}
  \BibitemShut {NoStop}%
\bibitem [{\citenamefont {Kobayashi}\ \emph {et~al.}(2012)\citenamefont
  {Kobayashi}, \citenamefont {Motohashi},\ and\ \citenamefont
  {Suyama}}]{Kobayashi:2012kh}%
  \BibitemOpen
  \bibfield  {author} {\bibinfo {author} {\bibfnamefont {T.}~\bibnamefont
  {Kobayashi}}, \bibinfo {author} {\bibfnamefont {H.}~\bibnamefont
  {Motohashi}}, \ and\ \bibinfo {author} {\bibfnamefont {T.}~\bibnamefont
  {Suyama}},\ }\href {\doibase 10.1103/PhysRevD.85.084025} {\bibfield
  {journal} {\bibinfo  {journal} {Phys. Rev.}\ }\textbf {\bibinfo {volume}
  {D85}},\ \bibinfo {pages} {084025} (\bibinfo {year} {2012})},\ \Eprint
  {http://arxiv.org/abs/1202.4893} {arXiv:1202.4893 [gr-qc]} \BibitemShut
  {NoStop}%
\bibitem [{\citenamefont {Kobayashi}\ \emph {et~al.}(2014)\citenamefont
  {Kobayashi}, \citenamefont {Motohashi},\ and\ \citenamefont
  {Suyama}}]{Kobayashi:2014wsa}%
  \BibitemOpen
  \bibfield  {author} {\bibinfo {author} {\bibfnamefont {T.}~\bibnamefont
  {Kobayashi}}, \bibinfo {author} {\bibfnamefont {H.}~\bibnamefont
  {Motohashi}}, \ and\ \bibinfo {author} {\bibfnamefont {T.}~\bibnamefont
  {Suyama}},\ }\href {\doibase 10.1103/PhysRevD.89.084042} {\bibfield
  {journal} {\bibinfo  {journal} {Phys. Rev.}\ }\textbf {\bibinfo {volume}
  {D89}},\ \bibinfo {pages} {084042} (\bibinfo {year} {2014})},\ \Eprint
  {http://arxiv.org/abs/1402.6740} {arXiv:1402.6740 [gr-qc]} \BibitemShut
  {NoStop}%
\bibitem [{\citenamefont {De~Felice}\ \emph {et~al.}(2011)\citenamefont
  {De~Felice}, \citenamefont {Suyama},\ and\ \citenamefont
  {Tanaka}}]{DeFelice:2011ka}%
  \BibitemOpen
  \bibfield  {author} {\bibinfo {author} {\bibfnamefont {A.}~\bibnamefont
  {De~Felice}}, \bibinfo {author} {\bibfnamefont {T.}~\bibnamefont {Suyama}}, \
  and\ \bibinfo {author} {\bibfnamefont {T.}~\bibnamefont {Tanaka}},\ }\href
  {\doibase 10.1103/PhysRevD.83.104035} {\bibfield  {journal} {\bibinfo
  {journal} {Phys. Rev.}\ }\textbf {\bibinfo {volume} {D83}},\ \bibinfo {pages}
  {104035} (\bibinfo {year} {2011})},\ \Eprint {http://arxiv.org/abs/1102.1521}
  {arXiv:1102.1521 [gr-qc]} \BibitemShut {NoStop}%
\bibitem [{\citenamefont {Motohashi}\ and\ \citenamefont
  {Suyama}(2011)}]{Motohashi:2011pw}%
  \BibitemOpen
  \bibfield  {author} {\bibinfo {author} {\bibfnamefont {H.}~\bibnamefont
  {Motohashi}}\ and\ \bibinfo {author} {\bibfnamefont {T.}~\bibnamefont
  {Suyama}},\ }\href {\doibase 10.1103/PhysRevD.84.084041} {\bibfield
  {journal} {\bibinfo  {journal} {Phys. Rev.}\ }\textbf {\bibinfo {volume}
  {D84}},\ \bibinfo {pages} {084041} (\bibinfo {year} {2011})},\ \Eprint
  {http://arxiv.org/abs/1107.3705} {arXiv:1107.3705 [gr-qc]} \BibitemShut
  {NoStop}%
\bibitem [{\citenamefont {Motohashi}\ and\ \citenamefont
  {Suyama}(2012)}]{Motohashi:2011ds}%
  \BibitemOpen
  \bibfield  {author} {\bibinfo {author} {\bibfnamefont {H.}~\bibnamefont
  {Motohashi}}\ and\ \bibinfo {author} {\bibfnamefont {T.}~\bibnamefont
  {Suyama}},\ }\href {\doibase 10.1103/PhysRevD.85.044054} {\bibfield
  {journal} {\bibinfo  {journal} {Phys. Rev.}\ }\textbf {\bibinfo {volume}
  {D85}},\ \bibinfo {pages} {044054} (\bibinfo {year} {2012})},\ \Eprint
  {http://arxiv.org/abs/1110.6241} {arXiv:1110.6241 [gr-qc]} \BibitemShut
  {NoStop}%
\bibitem [{\citenamefont {Ogawa}\ \emph {et~al.}(2016)\citenamefont {Ogawa},
  \citenamefont {Kobayashi},\ and\ \citenamefont {Suyama}}]{Ogawa:2015pea}%
  \BibitemOpen
  \bibfield  {author} {\bibinfo {author} {\bibfnamefont {H.}~\bibnamefont
  {Ogawa}}, \bibinfo {author} {\bibfnamefont {T.}~\bibnamefont {Kobayashi}}, \
  and\ \bibinfo {author} {\bibfnamefont {T.}~\bibnamefont {Suyama}},\ }\href
  {\doibase 10.1103/PhysRevD.93.064078} {\bibfield  {journal} {\bibinfo
  {journal} {Phys. Rev.}\ }\textbf {\bibinfo {volume} {D93}},\ \bibinfo {pages}
  {064078} (\bibinfo {year} {2016})},\ \Eprint
  {http://arxiv.org/abs/1510.07400} {arXiv:1510.07400 [gr-qc]} \BibitemShut
  {NoStop}%
\bibitem [{\citenamefont {Takahashi}\ \emph {et~al.}(2016)\citenamefont
  {Takahashi}, \citenamefont {Suyama},\ and\ \citenamefont
  {Kobayashi}}]{Takahashi:2015pad}%
  \BibitemOpen
  \bibfield  {author} {\bibinfo {author} {\bibfnamefont {K.}~\bibnamefont
  {Takahashi}}, \bibinfo {author} {\bibfnamefont {T.}~\bibnamefont {Suyama}}, \
  and\ \bibinfo {author} {\bibfnamefont {T.}~\bibnamefont {Kobayashi}},\ }\href
  {\doibase 10.1103/PhysRevD.93.064068} {\bibfield  {journal} {\bibinfo
  {journal} {Phys. Rev.}\ }\textbf {\bibinfo {volume} {D93}},\ \bibinfo {pages}
  {064068} (\bibinfo {year} {2016})},\ \Eprint
  {http://arxiv.org/abs/1511.06083} {arXiv:1511.06083 [gr-qc]} \BibitemShut
  {NoStop}%
\bibitem [{\citenamefont {Takahashi}\ and\ \citenamefont
  {Suyama}(2017)}]{Takahashi:2016dnv}%
  \BibitemOpen
  \bibfield  {author} {\bibinfo {author} {\bibfnamefont {K.}~\bibnamefont
  {Takahashi}}\ and\ \bibinfo {author} {\bibfnamefont {T.}~\bibnamefont
  {Suyama}},\ }\href {\doibase 10.1103/PhysRevD.95.024034} {\bibfield
  {journal} {\bibinfo  {journal} {Phys. Rev.}\ }\textbf {\bibinfo {volume}
  {D95}},\ \bibinfo {pages} {024034} (\bibinfo {year} {2017})},\ \Eprint
  {http://arxiv.org/abs/1610.00432} {arXiv:1610.00432 [gr-qc]} \BibitemShut
  {NoStop}%
\bibitem [{\citenamefont {Kase}\ \emph {et~al.}(2014)\citenamefont {Kase},
  \citenamefont {Gergely},\ and\ \citenamefont {Tsujikawa}}]{Kase:2014baa}%
  \BibitemOpen
  \bibfield  {author} {\bibinfo {author} {\bibfnamefont {R.}~\bibnamefont
  {Kase}}, \bibinfo {author} {\bibfnamefont {L.~A.}\ \bibnamefont {Gergely}}, \
  and\ \bibinfo {author} {\bibfnamefont {S.}~\bibnamefont {Tsujikawa}},\ }\href
  {\doibase 10.1103/PhysRevD.90.124019} {\bibfield  {journal} {\bibinfo
  {journal} {Phys. Rev.}\ }\textbf {\bibinfo {volume} {D90}},\ \bibinfo {pages}
  {124019} (\bibinfo {year} {2014})},\ \Eprint {http://arxiv.org/abs/1406.2402}
  {arXiv:1406.2402 [hep-th]} \BibitemShut {NoStop}%
\bibitem [{\citenamefont {Unruh}(1976)}]{PhysRevD.14.870}%
  \BibitemOpen
  \bibfield  {author} {\bibinfo {author} {\bibfnamefont {W.~G.}\ \bibnamefont
  {Unruh}},\ }\href {\doibase 10.1103/PhysRevD.14.870} {\bibfield  {journal}
  {\bibinfo  {journal} {Phys. Rev. D}\ }\textbf {\bibinfo {volume} {14}},\
  \bibinfo {pages} {870} (\bibinfo {year} {1976})}\BibitemShut {NoStop}%
\bibitem [{\citenamefont {{Harvey}}\ and\ \citenamefont
  {{Strominger}}(1993)}]{1993stqg.conf..122H}%
  \BibitemOpen
  \bibfield  {author} {\bibinfo {author} {\bibfnamefont {J.~A.}\ \bibnamefont
  {{Harvey}}}\ and\ \bibinfo {author} {\bibfnamefont {A.}~\bibnamefont
  {{Strominger}}},\ }in\ \href@noop {} {\emph {\bibinfo {booktitle} {String
  Theory and Quantum Gravity '92}}},\ \bibinfo {editor} {edited by\ \bibinfo
  {editor} {\bibfnamefont {J.}~\bibnamefont {{Harvey}}}, \bibinfo {editor}
  {\bibfnamefont {R.}~\bibnamefont {{Iengo}}}, \bibinfo {editor} {\bibfnamefont
  {K.~S.}\ \bibnamefont {{Narain}}}, \bibinfo {editor} {\bibfnamefont
  {S.}~\bibnamefont {{Randjbar-Daemi}}}, \ and\ \bibinfo {editor}
  {\bibfnamefont {H.}~\bibnamefont {{Verlinde}}}}\ (\bibinfo {year} {1993})\
  p.\ \bibinfo {pages} {122},\ \Eprint {http://arxiv.org/abs/hep-th/9209055}
  {hep-th/9209055} \BibitemShut {NoStop}%
\bibitem [{\citenamefont {Vagenas}(2002)}]{Vagenas:2001sm}%
  \BibitemOpen
  \bibfield  {author} {\bibinfo {author} {\bibfnamefont {E.~C.}\ \bibnamefont
  {Vagenas}},\ }\href {\doibase 10.1142/S0217732302006862} {\bibfield
  {journal} {\bibinfo  {journal} {Mod. Phys. Lett.}\ }\textbf {\bibinfo
  {volume} {A17}},\ \bibinfo {pages} {609} (\bibinfo {year} {2002})},\ \Eprint
  {http://arxiv.org/abs/hep-th/0108147} {arXiv:hep-th/0108147 [hep-th]}
  \BibitemShut {NoStop}%
\bibitem [{\citenamefont {Misner}\ \emph {et~al.}(1973)\citenamefont {Misner},
  \citenamefont {Thorne},\ and\ \citenamefont
  {Wheeler}}]{misner1973gravitation}%
  \BibitemOpen
  \bibfield  {author} {\bibinfo {author} {\bibfnamefont {C.~W.}\ \bibnamefont
  {Misner}}, \bibinfo {author} {\bibfnamefont {K.~S.}\ \bibnamefont {Thorne}},
  \ and\ \bibinfo {author} {\bibfnamefont {J.~A.}\ \bibnamefont {Wheeler}},\
  }\href@noop {} {\emph {\bibinfo {title} {Gravitation}}}\ (\bibinfo
  {publisher} {Macmillan},\ \bibinfo {year} {1973})\BibitemShut {NoStop}%
\bibitem [{\citenamefont {Brady}\ \emph {et~al.}(1996)\citenamefont {Brady},
  \citenamefont {Droz}, \citenamefont {Israel},\ and\ \citenamefont
  {Morsink}}]{Brady:1995na}%
  \BibitemOpen
  \bibfield  {author} {\bibinfo {author} {\bibfnamefont {P.~R.}\ \bibnamefont
  {Brady}}, \bibinfo {author} {\bibfnamefont {S.}~\bibnamefont {Droz}},
  \bibinfo {author} {\bibfnamefont {W.}~\bibnamefont {Israel}}, \ and\ \bibinfo
  {author} {\bibfnamefont {S.~M.}\ \bibnamefont {Morsink}},\ }\href {\doibase
  10.1088/0264-9381/13/8/015} {\bibfield  {journal} {\bibinfo  {journal}
  {Class. Quant. Grav.}\ }\textbf {\bibinfo {volume} {13}},\ \bibinfo {pages}
  {2211} (\bibinfo {year} {1996})},\ \Eprint
  {http://arxiv.org/abs/gr-qc/9510040} {arXiv:gr-qc/9510040 [gr-qc]}
  \BibitemShut {NoStop}%
\bibitem [{\citenamefont {Bardeen}(1980)}]{PhysRevD.22.1882}%
  \BibitemOpen
  \bibfield  {author} {\bibinfo {author} {\bibfnamefont {J.~M.}\ \bibnamefont
  {Bardeen}},\ }\href {\doibase 10.1103/PhysRevD.22.1882} {\bibfield  {journal}
  {\bibinfo  {journal} {Phys. Rev. D}\ }\textbf {\bibinfo {volume} {22}},\
  \bibinfo {pages} {1882} (\bibinfo {year} {1980})}\BibitemShut {NoStop}%
\bibitem [{\citenamefont {Motohashi}\ \emph {et~al.}(2016)\citenamefont
  {Motohashi}, \citenamefont {Suyama},\ and\ \citenamefont
  {Takahashi}}]{Motohashi:2016prk}%
  \BibitemOpen
  \bibfield  {author} {\bibinfo {author} {\bibfnamefont {H.}~\bibnamefont
  {Motohashi}}, \bibinfo {author} {\bibfnamefont {T.}~\bibnamefont {Suyama}}, \
  and\ \bibinfo {author} {\bibfnamefont {K.}~\bibnamefont {Takahashi}},\ }\href
  {\doibase 10.1103/PhysRevD.94.124021} {\bibfield  {journal} {\bibinfo
  {journal} {Phys. Rev.}\ }\textbf {\bibinfo {volume} {D94}},\ \bibinfo {pages}
  {124021} (\bibinfo {year} {2016})},\ \Eprint
  {http://arxiv.org/abs/1608.00071} {arXiv:1608.00071 [gr-qc]} \BibitemShut
  {NoStop}%
\bibitem [{\citenamefont {Frolov}\ and\ \citenamefont
  {Novikov}(2012)}]{frolov2012black}%
  \BibitemOpen
  \bibfield  {author} {\bibinfo {author} {\bibfnamefont {V.}~\bibnamefont
  {Frolov}}\ and\ \bibinfo {author} {\bibfnamefont {I.}~\bibnamefont
  {Novikov}},\ }\href@noop {} {\emph {\bibinfo {title} {Black hole physics:
  basic concepts and new developments}}},\ Vol.~\bibinfo {volume} {96}\
  (\bibinfo  {publisher} {Springer Science \& Business Media},\ \bibinfo {year}
  {2012})\BibitemShut {NoStop}%
\bibitem [{\citenamefont {Sarbach}\ and\ \citenamefont
  {Tiglio}(2001)}]{Sarbach:2001qq}%
  \BibitemOpen
  \bibfield  {author} {\bibinfo {author} {\bibfnamefont {O.}~\bibnamefont
  {Sarbach}}\ and\ \bibinfo {author} {\bibfnamefont {M.}~\bibnamefont
  {Tiglio}},\ }\href {\doibase 10.1103/PhysRevD.64.084016} {\bibfield
  {journal} {\bibinfo  {journal} {Phys. Rev.}\ }\textbf {\bibinfo {volume}
  {D64}},\ \bibinfo {pages} {084016} (\bibinfo {year} {2001})},\ \Eprint
  {http://arxiv.org/abs/gr-qc/0104061} {arXiv:gr-qc/0104061 [gr-qc]}
  \BibitemShut {NoStop}%
\bibitem [{\citenamefont {Sarbach}(2000)}]{SarbachThesis}%
  \BibitemOpen
  \bibfield  {author} {\bibinfo {author} {\bibfnamefont {O.}~\bibnamefont
  {Sarbach}},\ }\emph {\bibinfo {title} {On the generalization of the
  Regge-Wheeler equation for self-gravitating matter fields}},\ \href@noop {}
  {Ph.D. thesis},\ \bibinfo  {school} {Universit\"{a}t Z\"{u}rich} (\bibinfo
  {year} {2000})\BibitemShut {NoStop}%
\bibitem [{\citenamefont {Martel}(2003)}]{MartelThesis}%
  \BibitemOpen
  \bibfield  {author} {\bibinfo {author} {\bibfnamefont {K.}~\bibnamefont
  {Martel}},\ }\emph {\bibinfo {title} {Particles and black holes: time-domain
  integration of the equations of the equations of black-hole perturbation
  theory}},\ \href@noop {} {Ph.D. thesis},\ \bibinfo  {school} {The University
  of Guelph} (\bibinfo {year} {2003})\BibitemShut {NoStop}%
\bibitem [{\citenamefont {Fierz}\ and\ \citenamefont
  {Pauli}(1939)}]{Fierz:1939ix}%
  \BibitemOpen
  \bibfield  {author} {\bibinfo {author} {\bibfnamefont {M.}~\bibnamefont
  {Fierz}}\ and\ \bibinfo {author} {\bibfnamefont {W.}~\bibnamefont {Pauli}},\
  }\href {\doibase 10.1098/rspa.1939.0140} {\bibfield  {journal} {\bibinfo
  {journal} {Proc. Roy. Soc. Lond.}\ }\textbf {\bibinfo {volume} {A173}},\
  \bibinfo {pages} {211} (\bibinfo {year} {1939})}\BibitemShut {NoStop}%
\bibitem [{\citenamefont {de~Rham}\ \emph {et~al.}(2011)\citenamefont
  {de~Rham}, \citenamefont {Gabadadze},\ and\ \citenamefont
  {Tolley}}]{deRham:2010kj}%
  \BibitemOpen
  \bibfield  {author} {\bibinfo {author} {\bibfnamefont {C.}~\bibnamefont
  {de~Rham}}, \bibinfo {author} {\bibfnamefont {G.}~\bibnamefont {Gabadadze}},
  \ and\ \bibinfo {author} {\bibfnamefont {A.~J.}\ \bibnamefont {Tolley}},\
  }\href {\doibase 10.1103/PhysRevLett.106.231101} {\bibfield  {journal}
  {\bibinfo  {journal} {Phys. Rev. Lett.}\ }\textbf {\bibinfo {volume} {106}},\
  \bibinfo {pages} {231101} (\bibinfo {year} {2011})},\ \Eprint
  {http://arxiv.org/abs/1011.1232} {arXiv:1011.1232 [hep-th]} \BibitemShut
  {NoStop}%
\bibitem [{\citenamefont {Hinterbichler}(2012)}]{Hinterbichler:2011tt}%
  \BibitemOpen
  \bibfield  {author} {\bibinfo {author} {\bibfnamefont {K.}~\bibnamefont
  {Hinterbichler}},\ }\href {\doibase 10.1103/RevModPhys.84.671} {\bibfield
  {journal} {\bibinfo  {journal} {Rev. Mod. Phys.}\ }\textbf {\bibinfo {volume}
  {84}},\ \bibinfo {pages} {671} (\bibinfo {year} {2012})},\ \Eprint
  {http://arxiv.org/abs/1105.3735} {arXiv:1105.3735 [hep-th]} \BibitemShut
  {NoStop}%
\bibitem [{\citenamefont {de~Rham}(2014)}]{deRham:2014zqa}%
  \BibitemOpen
  \bibfield  {author} {\bibinfo {author} {\bibfnamefont {C.}~\bibnamefont
  {de~Rham}},\ }\href {\doibase 10.12942/lrr-2014-7} {\bibfield  {journal}
  {\bibinfo  {journal} {Living Rev. Rel.}\ }\textbf {\bibinfo {volume} {17}},\
  \bibinfo {pages} {7} (\bibinfo {year} {2014})},\ \Eprint
  {http://arxiv.org/abs/1401.4173} {arXiv:1401.4173 [hep-th]} \BibitemShut
  {NoStop}%
\bibitem [{\citenamefont {Newman}\ and\ \citenamefont
  {Penrose}(1962)}]{doi:10.1063/1.1724257}%
  \BibitemOpen
  \bibfield  {author} {\bibinfo {author} {\bibfnamefont {E.}~\bibnamefont
  {Newman}}\ and\ \bibinfo {author} {\bibfnamefont {R.}~\bibnamefont
  {Penrose}},\ }\href {\doibase 10.1063/1.1724257} {\bibfield  {journal}
  {\bibinfo  {journal} {Journal of Mathematical Physics}\ }\textbf {\bibinfo
  {volume} {3}},\ \bibinfo {pages} {566} (\bibinfo {year} {1962})}\BibitemShut
  {NoStop}%
\bibitem [{\citenamefont {Ray}(2008)}]{SouryaRayThesis2p2}%
  \BibitemOpen
  \bibfield  {author} {\bibinfo {author} {\bibfnamefont {S.}~\bibnamefont
  {Ray}},\ }\emph {\bibinfo {title} {Investigation of symmetries and conserved
  charges in general relativity}},\ \href@noop {} {Ph.D. thesis},\ \bibinfo
  {school} {University of Massachusetts-Amherst} (\bibinfo {year}
  {2008})\BibitemShut {NoStop}%
\bibitem [{\citenamefont {{Carter}}(1992)}]{1992JGP.....8...53C}%
  \BibitemOpen
  \bibfield  {author} {\bibinfo {author} {\bibfnamefont {B.}~\bibnamefont
  {{Carter}}},\ }\href {\doibase 10.1016/0393-0440(92)90043-Z} {\bibfield
  {journal} {\bibinfo  {journal} {Journal of Geometry and Physics}\ }\textbf
  {\bibinfo {volume} {8}},\ \bibinfo {pages} {53} (\bibinfo {year}
  {1992})}\BibitemShut {NoStop}%
\bibitem [{\citenamefont {Gourgoulhon}(2005)}]{Gourgoulhon:2005ch}%
  \BibitemOpen
  \bibfield  {author} {\bibinfo {author} {\bibfnamefont {E.}~\bibnamefont
  {Gourgoulhon}},\ }\href {\doibase 10.1103/PhysRevD.72.104007} {\bibfield
  {journal} {\bibinfo  {journal} {Phys. Rev.}\ }\textbf {\bibinfo {volume}
  {D72}},\ \bibinfo {pages} {104007} (\bibinfo {year} {2005})},\ \Eprint
  {http://arxiv.org/abs/gr-qc/0508003} {arXiv:gr-qc/0508003 [gr-qc]}
  \BibitemShut {NoStop}%
\bibitem [{\citenamefont {{Damour}}(1982)}]{1982mgm..conf..587D}%
  \BibitemOpen
  \bibfield  {author} {\bibinfo {author} {\bibfnamefont {T.}~\bibnamefont
  {{Damour}}},\ }in\ \href@noop {} {\emph {\bibinfo {booktitle} {Marcel
  Grossmann Meeting: General Relativity}}},\ \bibinfo {editor} {edited by\
  \bibinfo {editor} {\bibfnamefont {R.}~\bibnamefont {{Ruffini}}}}\ (\bibinfo
  {year} {1982})\BibitemShut {NoStop}%
\bibitem [{\citenamefont {Damour}(1978)}]{PhysRevD.18.3598}%
  \BibitemOpen
  \bibfield  {author} {\bibinfo {author} {\bibfnamefont {T.}~\bibnamefont
  {Damour}},\ }\href {\doibase 10.1103/PhysRevD.18.3598} {\bibfield  {journal}
  {\bibinfo  {journal} {Phys. Rev. D}\ }\textbf {\bibinfo {volume} {18}},\
  \bibinfo {pages} {3598} (\bibinfo {year} {1978})}\BibitemShut {NoStop}%
\bibitem [{\citenamefont {Damour}(1979)}]{DamourThesis}%
  \BibitemOpen
  \bibfield  {author} {\bibinfo {author} {\bibfnamefont {T.}~\bibnamefont
  {Damour}},\ }\emph {\bibinfo {title} {Quelues propri\'{e}t\'{e}s
  m\'{e}caniques, \'{e}lectromagn\'{e}tiques, thermodynamiques et quantiques
  des trous noirs}},\ \href@noop {} {Ph.D. thesis},\ \bibinfo  {school}
  {Universit\'{e} Pierre et Marie Curie, Paris 6} (\bibinfo {year}
  {1979})\BibitemShut {NoStop}%
\bibitem [{\citenamefont {Padmanabhan}(2011)}]{Padmanabhan:2010rp}%
  \BibitemOpen
  \bibfield  {author} {\bibinfo {author} {\bibfnamefont {T.}~\bibnamefont
  {Padmanabhan}},\ }\href {\doibase 10.1103/PhysRevD.83.044048} {\bibfield
  {journal} {\bibinfo  {journal} {Phys. Rev.}\ }\textbf {\bibinfo {volume}
  {D83}},\ \bibinfo {pages} {044048} (\bibinfo {year} {2011})},\ \Eprint
  {http://arxiv.org/abs/1012.0119} {arXiv:1012.0119 [gr-qc]} \BibitemShut
  {NoStop}%
\bibitem [{\citenamefont {Poisson}(2004)}]{PoissonBook}%
  \BibitemOpen
  \bibfield  {author} {\bibinfo {author} {\bibfnamefont {E.}~\bibnamefont
  {Poisson}},\ }\href@noop {} {\emph {\bibinfo {title} {A relativist's toolkit
  : the mathematics of black-hole mechanics}}}\ (\bibinfo  {publisher}
  {Cambridge University Press},\ \bibinfo {year} {2004})\BibitemShut {NoStop}%
\bibitem [{\citenamefont {{Hinterbichler}}\ \emph {et~al.}(2010)\citenamefont
  {{Hinterbichler}}, \citenamefont {{Trodden}},\ and\ \citenamefont
  {{Wesley}}}]{2010PhRvD..82l4018H}%
  \BibitemOpen
  \bibfield  {author} {\bibinfo {author} {\bibfnamefont {K.}~\bibnamefont
  {{Hinterbichler}}}, \bibinfo {author} {\bibfnamefont {M.}~\bibnamefont
  {{Trodden}}}, \ and\ \bibinfo {author} {\bibfnamefont {D.}~\bibnamefont
  {{Wesley}}},\ }\href {\doibase 10.1103/PhysRevD.82.124018} {\bibfield
  {journal} {\bibinfo  {journal} {Phys. Rev. D.}\ }\textbf {\bibinfo {volume}
  {82}},\ \bibinfo {eid} {124018} (\bibinfo {year} {2010})},\ \Eprint
  {http://arxiv.org/abs/1008.1305} {arXiv:1008.1305 [hep-th]} \BibitemShut
  {NoStop}%
\bibitem [{\citenamefont {Grant}\ and\ \citenamefont
  {Moss}(1997)}]{PhysRevD.56.6284}%
  \BibitemOpen
  \bibfield  {author} {\bibinfo {author} {\bibfnamefont {J.~D.~E.}\
  \bibnamefont {Grant}}\ and\ \bibinfo {author} {\bibfnamefont {I.~G.}\
  \bibnamefont {Moss}},\ }\href {\doibase 10.1103/PhysRevD.56.6284} {\bibfield
  {journal} {\bibinfo  {journal} {Phys. Rev. D}\ }\textbf {\bibinfo {volume}
  {56}},\ \bibinfo {pages} {6284} (\bibinfo {year} {1997})}\BibitemShut
  {NoStop}%
\end{thebibliography}%

\end{document}